\documentclass[journal=jacsat,manuscript=article]{achemso}
\usepackage{graphicx}
\usepackage{bm}
\usepackage{mathptmx}
\usepackage{epstopdf}
\usepackage{booktabs}
\usepackage{multirow}
\usepackage{hhline}
\usepackage{dcolumn}
\usepackage{subfigure}
\usepackage{amsmath}
\usepackage{float}
\usepackage{chemformula}
\usepackage[version=3]{mhchem}
\usepackage{chemfig}
\usepackage{soul}
 
\title
{Identifying the critical surface descriptors responsible for the appearance of negative slopes in the adsorption energy scaling relationships}

\author{Swetarekha Ram, Seung-Cheol Lee$^{\ast}$ and Satadeep Bhattacharjee}
\email{seungcheol.lee@ikst.res.in, s.bhattacharjee@ikst.res.in}

\affiliation{Indo-Korea Science and Technology Center (IKST),  Bangalore-560064, India}

\begin{document}
\begin{abstract}
Adsorption energy scaling relationships have now developed beyond their original form, which was more targeted towards the optimization of catalytic sites and the reduction of computational costs in simulations. The recent surge of interest in the adsorption energy scaling relations is to explore the surfaces beyond the transition metals (TMs) as well as reactions involving complicated molecules. Breakdown of such scaling relationships leads to motivating the discovery of novel catalysts with enhanced capabilities. In this work, we report our extensive study on the linear scaling relation (LSR) between oxygen (O), a group VIA element with elements of neighbouring groups such as: Group IIIA (Boron (B), Aluminum (Al)), IVA (Carbon (C), Silicon (Si)), VA (Nitrogen(N),phosphorus(P)) and VIIA (Florine(F)) on magnetic bimetallic surfaces. We found that the slope is positive for only O versus N and F, remaining of the slopes are negative. The present model is based on multiple surface descriptors, particularly spin-averaged d-band center and the surface's magnetic moment, whereas the original scaling theory (Phys. Rev. Lett. 99, 016105 (2007)) was based on a single adsorbate descriptor: adsorbate valency.



\end{abstract}

\maketitle

\section{Introduction}
In the last decades, the first-principles methods coupled with increasing computing power has enabled to generate large data sets for describing the interaction between adsorbates with different solid surfaces. This has aided in the conceptualization of energy trends through scaling relations. Much research has been devoted to the adsorption energy of hydrogenated atomic adsorbate on transition metals and hence towards scaling relation, due to their importance not only in the fundamental surface science but also in the petrochemical industry. In recent years researchers
have turned to developing catalysts using cheap and abundant 3d transition metals (TMs) and their alloys due to the low abundance and high cost of 4d and 5d metals. Magnetic bimetallic  transition metals are an interesting class of catalysts materials, on which the scaling relations are relatively unexplored. In this regard, the role of magnetism in heterogeneous catalysis is the subject of a recent study. \cite{chretien20082,mtangi2015role,torun2013role} The catalytic reactions in the transition metal (TM) surfaces can also be tuned using alloying. The development of reliable Pt-based electrocatalysts with a wide range of compositions can be possible by alloying Pt with TM, which reduces the usage of Pt and at the same time improving performance as compared with that of pure Pt.\cite{stamenkovic2007trends,stamenkovic2007improved,zhang2009general,
kang2010synthesis} One such common approach that has been used to explore the efficient design of Oxygen Reduction Reaction (ORR) catalysts is alloying Pt with the first row TM like Mn, Fe, Co or Ni.\cite{duan2011first,wang2013structurally} 
The difference in the interactions among the adsorbed species with heterogeneous catalysts dictates the relative activities of catalysts. \cite{stegelmann2009degree} In the past decade one of the leading accomplishments of theoretical surfaces science and in heterogeneous catalysis come upon scaling laws between the adsorption energies of atoms and their hydrogenated species on TM surface. \cite{abild2007scaling,norskov2009towards} However, seldom the possibility of scalability among atomic adsorbates, which are not bonded with hydrogen atoms was also claimed by Calle-Vallejo \textit{et. al}\cite{calle2012physical}. 

So far, studies based on adsorption properties are focused on non-magnetic TM surfaces. For the magnetic surfaces, information is limited. Furthermore, the scaling laws involving only atomic adsorbates are limited \cite{calle2012physical}. In fact, the correlation of adsorption energy of adsorbates through scaling laws is believed to limit the search of interesting catalysts. The fundamental understanding of the electronic structure and the adsorbate-surface interaction fascinates towards realization of the making and breaking of chemical bonds of molecules. In the field of catalysis research, the main objective is to design and tune the catalyst by controlling their structural properties. Thus, in recent times, a general curiosity for breaking scaling relation is a realistic approach for both theoretical and experimental studies. \cite{khorshidi2018strain,wang2017breaking, gani2018understanding,li2016recent} Indeed, a desirable goal is to obtain a negative slope in linear scaling relation (LSR) between the atomic adsorbates. Herein, we show the importance of the d-band center in appearance of the negative slope in linear scaling relation (LSR) of magnetic bimetallic TMs. During the last two
decades, the d-band model has been widely used to understand
variations in chemisorption energies of various adsorbates on
transition-metal surfaces and their alloys.\cite{mavrikakis1998effect,kitchin2004role,kitchin2004modification} To scrutinize the scaling
relations in the estimation of atomic adsorption energy on the magnetic bimetallic TM surfaces we perform an extensive theoretical investigation, based on quantum-mechanical first-principles calculations within the density-functional theory (DFT).

  In the present study, the specific interest is to deviate from the standard scaling relationship in order to achieve higher activity or selectivity in a catalytic reaction. Here we explore the limitation of universally accepted LSR that can be used to design new and screen for better heterogeneous catalysts.
               In order to understand the nature of chemisorption, we perform a systematic study of the adsorption on the magnetic bimetallic  surface for atomic adsorbates considering the second row elements in the periodic table viz. B, C, N, F and O and third row elements in the periodic table viz. Al, Si, and P. Through the detailed analysis of the electronic structure upon adsorption, the nature of chemisorption is realized and is presented. We investigate the cause for the different natures of slope for atomic adsorbates in scaling behavior and discuss implications to the prediction of the best catalytic material. Prior to this, in our earlier study, deviation of standard LSR for O, C, N and their hydrogenated species: OH, CH$_x$, NH$_x$ ($x=1,2$), respectively on the magnetic bimetallic TMs has been noticed \cite{ram2020adsorption}.

\section{Computational details} 
 Our theoretical estimates of optimized lattice parameters of $L1_{0}$ phase of NiPt, FePt, FePd, MnPt, MnPd,
CoPt and L1$_2$ phase of CoPt$_3$, MnPt$_3$, FePt$_3$, NiPt$_3$ and Co$_3$Pt of bulk  magnetic bimetallic
compounds are chosen
to model
a periodically
repeating 8-layer  (111) slab in our calculations (see the optimized lattice parameters and magnetic moment in bulk form of selected compounds in Table-S1 in supplementary information (SI)). We determined electronic properties using first-principles spin-polarized density functional theory (DFT) based calculations, with the projector-augmented wave (PAW) method as implemented in the Vienna Ab Initio Simulation Package (VASP).\cite{kresse1996efficient, kresse1996efficiency} 
We treated exchange-correlation energy of electrons within the generalized gradient approximation (GGA) with the Perdew-Burke-Ernzerhof (PBE)\cite{perdew1996rationale, perdew1996phys} functional form. 
In all the calculations, we have used ultrasoft pseudopotentials\cite{kresse1999ultrasoft} 
and a kinetic energy cutoff of 500 eV to expand the electronic
wave functions in the plane-wave basis set.  For the energy (stress) convergence the criterion of $10^{-6}$ eV is applied. 
 The bulk lattice constants of all magnetic bimetallic TMs  are optimized using $12 \times 12 \times 12$ Monkhorst-Pack\cite{monkhorst1976special} type of $k$-point sampling. $12 \times 12 \times 1$ $k$-points are used for the Brillouin zone sampling in the case of slabs. To avoid stress in the adsorption energy calculations, the lower four layers are constrained to the DFT-calculated bulk geometry, while atoms in the upper four layers and adatoms are
unconstrained. The vacuum region between the slabs normal to the
surface is kept as 20 $\AA$ in order to have the negligible interaction between the periodic images. Adsorption energies are calculated
for all adsorption sites in (111) surfaces like fcc, hcp, bridge,
and on-top site. For calculating LSR, we
have chosen the adsorption energy at the most favorable site (the calculated adsorption energy of selected adatoms are in Table-S2 in SI).

A dipole correction is used to ensure that there is no unwanted electric field arising from the asymmetry of the cell along the c-direction.
Despite the fact that
the magnetic  bimetallic TM surfaces do not contain intrinsic
dipoles, the adatom on one side of the slab
gives rise to a dipole perpendicular to the surface. Therefore we have
 employed the dipole correction to nullify the artificial
field imposed on the slab by the periodic boundary conditions to
obtain the correct adsorption energies\cite{bengtsson1999dipole}.

The adsorption energy ($\Delta E_{ads} $) of adsorbate (A) are calculated relative to their clean surface and the isolated atoms, as

\begin{equation}
\Delta E_{ads} = E_{surface + A} - E_{surface} - E_{A}
\label{equ:ads}
\end{equation}

Where $ E_{surface + A} $ denotes the energy of adsorbed active site of the surface, $  E_{surface} $ and $ E_{A} $, represent the energy of the clean surface and the energy of the isolated atom, respectively.

\section{Results and discussions}

\subsection{Linear relation of adsorption energy}

The catalytic activity between the given pair of species 1 and 2 generally mapped through the scaling relations of adsorption energy and is expressed as  

\begin{equation}
\Delta E_{ads}^{1} = \gamma \Delta E_{ads}^{2} +\xi
\label{equ:scaling}
\end{equation}

Where $\Delta E_{ads}^{1} $ and $\Delta E_{ads}^{2} $ are the adsorption energy of species 1 and 2 on  magnetic bimetallic TM (111) surface. It is believed that the slope $\gamma$ on a close-packed surface depends on the ratio of the number of valence electrons of the adatoms that bound to the surface\cite{calle2015introducing,calle2012physical} or it is the ratio of the bond orders\cite{abild2016computational,van2016molecular}  and the intercept $\xi$ depends on surface coordination\cite{calle2015introducing}. In the present study,  $\xi$ is constant as we have selected a close-packed surface for all magnetic bimetallic TMs. The nature of the slope in the case of standard scaling relation on TM surfaces as available in the literature \cite{tsai2014understanding,calle2015introducing,fields2017scaling} limit the catalytic activity in the field of surface science. 
Therefore it is interesting to explore the adsorption energy of the atomic adsorbates on magnetic bimetallic TM close-packed surface to check whether the slope of the LSR can be modified to select the best catalytic material.

\begin{figure}[]
  \centering
     {\includegraphics[scale=0.5]{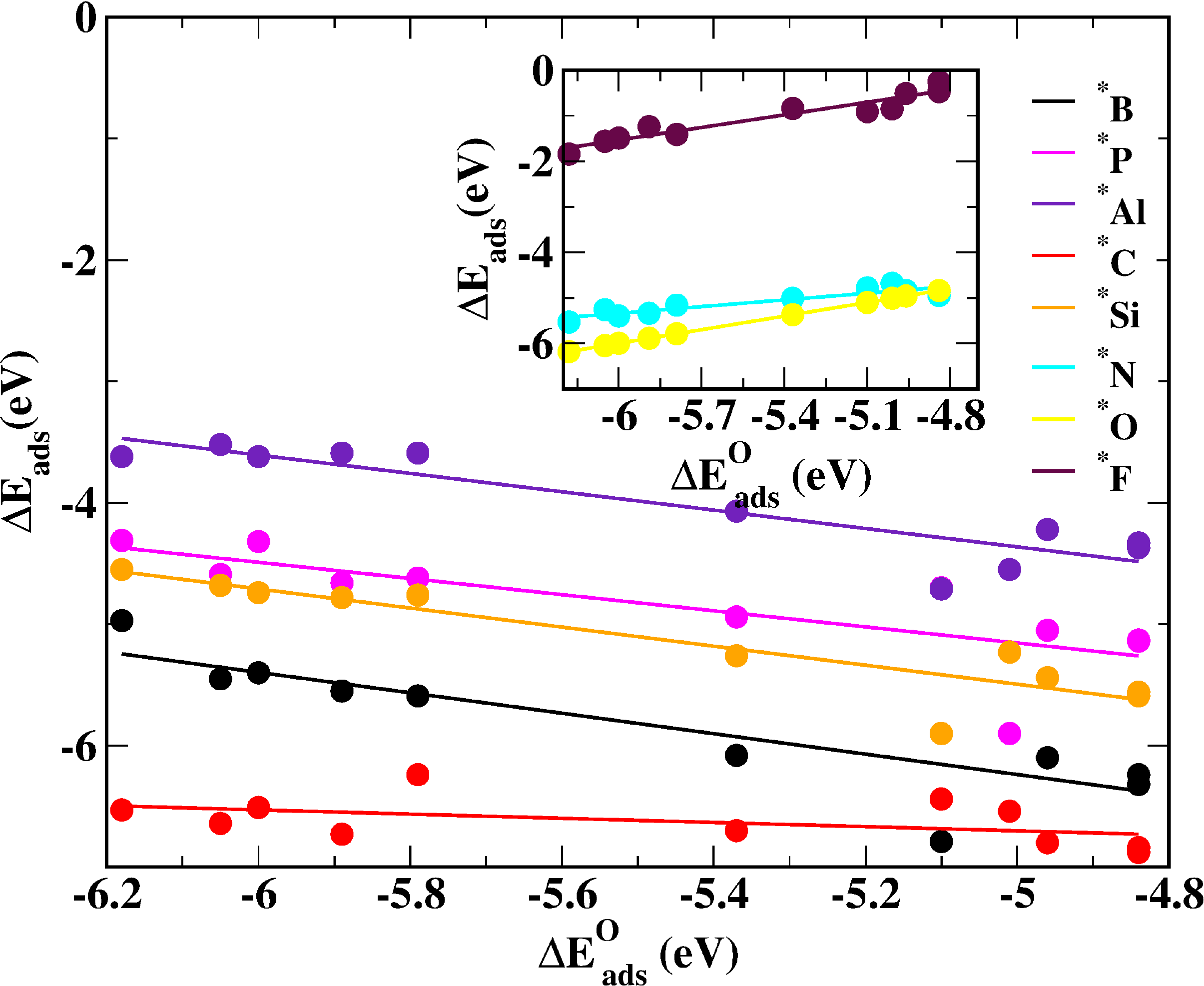}}    
 \caption{Scaling relations for the adsorption energies $\Delta E_{ads}$ of $ O $ and $ 2p $  and  $ 3p $ atoms on magnetic bimetallic TM surfaces. Inset shows the positive slope only for  $\Delta E^{O}_{ads}$ vs   $\Delta E^{N}_{ads}$ and $\Delta E^{F}_{ads}$ while for other $2p$ and $3p$ atoms negative slopes are evident.}
 \label{Fig:scalingOvs2p}
\end{figure}

Figure-\ref{Fig:scalingOvs2p} represents the linear correlation between the adsorption energies of the most stable site of adsorbed O and other atomic adsorbates belong to the second and third rows of the periodic table (B, C, N, F, Al, Si, P) on  magnetic bimetallic TM surfaces. So far the scaling relation of adsorption energy  between atomic adsorbates and their hydrogenated species were reported on TM surfaces, TM Oxide, Sulfide, and Nitride Surfaces, layered TM sulphides and doped molybdenum phosphide surfaces \cite{tsai2014understanding,fernandez2008scaling,calle2015introducing,fields2017scaling}. 
The existence of scaling relation between second and third row atoms of the periodic table was available on atop
sites of NSAs (near-surface
alloys) of Pt(111) and 3d, 4d, and 5d TMs\cite{calle2012physical}. Earlier, the appearance of negative slope in case of B on TM surface was presented by \textit{Hai-Yan et. al}\cite{su2016establishing}. However, in this study the presented result in Figure-\ref{Fig:scalingOvs2p} disclose the appearance of negative slope between $  O$ and other selected atomic adsorbates such as  B,  C,  P,  Al,  Si on  magnetic bimetallic TMs surface, whereas the slope is noticed to be positive in between $  O$ and $  N$, $  F$ . The value of the $ \gamma $ and $ \xi $ of scaling relation as in equation-\ref{equ:scaling} are presented in Table-\ref{tab-scaling}. This shows that binding of the atomic adsorbate to the surface directly determined by the binding of O; the stronger(weaker) O-magnetic bimetallic TM binding will have a weaker(stronger) binding with other atomic adsorbates except for N and F.

Eventually, appearance of negative slope in the LSR can be explained through the analysis of surface properties and adsorption energy correlation. The LSR between the adsorption energy of $    O$ with other $2p$ and $3p$ atoms indicates that the parameter which describes the $\Delta E^O_{ads}$ should also describe the $\Delta E^{(2p/3p)^{atoms}}_{ads}$.  The evident provided in Figure-\ref{Fig:scalingOvs2p} shows the presence of negative slope in between  O and other selected $2p$ and $3p$ atoms except in  N and  F. 
At the same time the slope of LSR between two adsorbates is the ratio of the number of valence electrons of the adsorbate as reported earlier\cite{calle2015introducing,calle2012physical} is not satisfied in the present case. To illustrates this we have extended the analysis for better understanding of nature of the surface and adsorbates.

\subsection{Connection between surface properties and adsorption energy scaling relationships}
\noindent 

In order to address the discrepancy in the usual LSR for the adsorption energy of $    O$ with other $2p$ and $3p$ atoms on  magnetic bimetallic TM surface we first examine the factors that basically play a vital role in the adsorption of atomic adsorbates on magnetic bimetallic TM surface. We propose the surface properties like the number of average valence electrons ($NV_{av}^{slab}$), magnetic moment of the surface ($m_{surf}$), work function ($\phi_{slab}$) of the surface and the $d$ band center of the surface ($\epsilon_d^{slab}$)  as active parameters which are decisive factor in the estimation of adsorption energy scaling relation. The $d$-band center of the surface was calculated by using spin-averaged d-band center, $\varepsilon_d$ method\cite{bhattacharjee2016improved} from equation-\ref{equ:dcenter}.

\begin{equation}
\varepsilon_d = \displaystyle {\sum_\sigma \frac{f_\sigma \varepsilon_{d\sigma}}{\displaystyle\sum_{\sigma} f_\sigma }}   
\label{equ:dcenter}
\end{equation}

Here  $f_\sigma$ is the spin-
dependent fractional occupations for the d-band and $\varepsilon_{d\sigma}$ is the $d$-band center for spin $\sigma$. $f_\sigma$ and $\varepsilon_{d\sigma}$ are calculated using equation- \ref{equ:dcenter2}.  

 \begin{equation}
 f_{\sigma} = \frac{\displaystyle\int\limits_{-\infty}^{E_F} D_{d\sigma}(E) dE} {5} \quad 
 and \quad
 \varepsilon_{d\sigma} =  \frac{\displaystyle\int\limits_{-\infty}^{\infty} ED_{d\sigma}(E-E_F) dE }{\displaystyle\int\limits_{-\infty}^{\infty} D_{d\sigma}(E-E_F) dE} 
 \label{equ:dcenter2}
\end{equation}

  $D_{d\sigma}$ is the projected density of states (DOS) on $d$-states of TMs for spin $\sigma$ in function of state energy E.  All 5 $d$ orbitals and the 3$d$ bands of two constituent TMs of magnetic bimetallic are treated for calculating $d$-band center.

Let us consider the adsorption energies of species 1 and 2 depend upon a set of surface variables, $\lbrace\omega_i\rbrace$, which further correlate the $\Delta E_{ads}^{1}$ and $\Delta E_{ads}^{2} $ as\cite{calle2012physical,calle2015introducing} 

\begin{equation}
\Delta E_{ads}^{1} = F(\lbrace\omega_i\rbrace) + \alpha_{0} \quad
 and \quad
\Delta E_{ads}^{2} = G(\lbrace\omega_i\rbrace) + \beta_{0}
\label{equ:function}
\end{equation}

where $F$ and $G$ are two functions of the set $\lbrace\omega_i\rbrace$ of the surface properties and $\alpha_{0} $ and $\beta_{0}$ depends on surface coordination number \cite{calle2015introducing,calle2014fast}, which are constant in the present case as all considered surface for adsorption are close-packed surface. At the same time the equation-\ref{equ:scaling}  
on  magnetic bimetallic TMs for atomic adsorbates  has to satisfy the following additional constrains given by the equation-\ref{equ:scalingn} and equation-\ref{equ:offset} as below \cite{calle2012physical,calle2015introducing} 


\begin{equation}
F (\lbrace\omega_i\rbrace) = \gamma G (\lbrace\omega_i\rbrace)\
\label{equ:scalingn}
\end{equation}
 Again the offset $\xi$ in equation-\ref{equ:scaling} links the  $\alpha_{0} $ and $\beta_{0}$ through the slope, $\gamma$ as

\begin{equation}
\xi = \alpha_0 -\gamma\beta_0
\label{equ:offset}
\end{equation}

\begin{figure}[]
  \centering
     \subfigure[]{\includegraphics[scale=0.3]{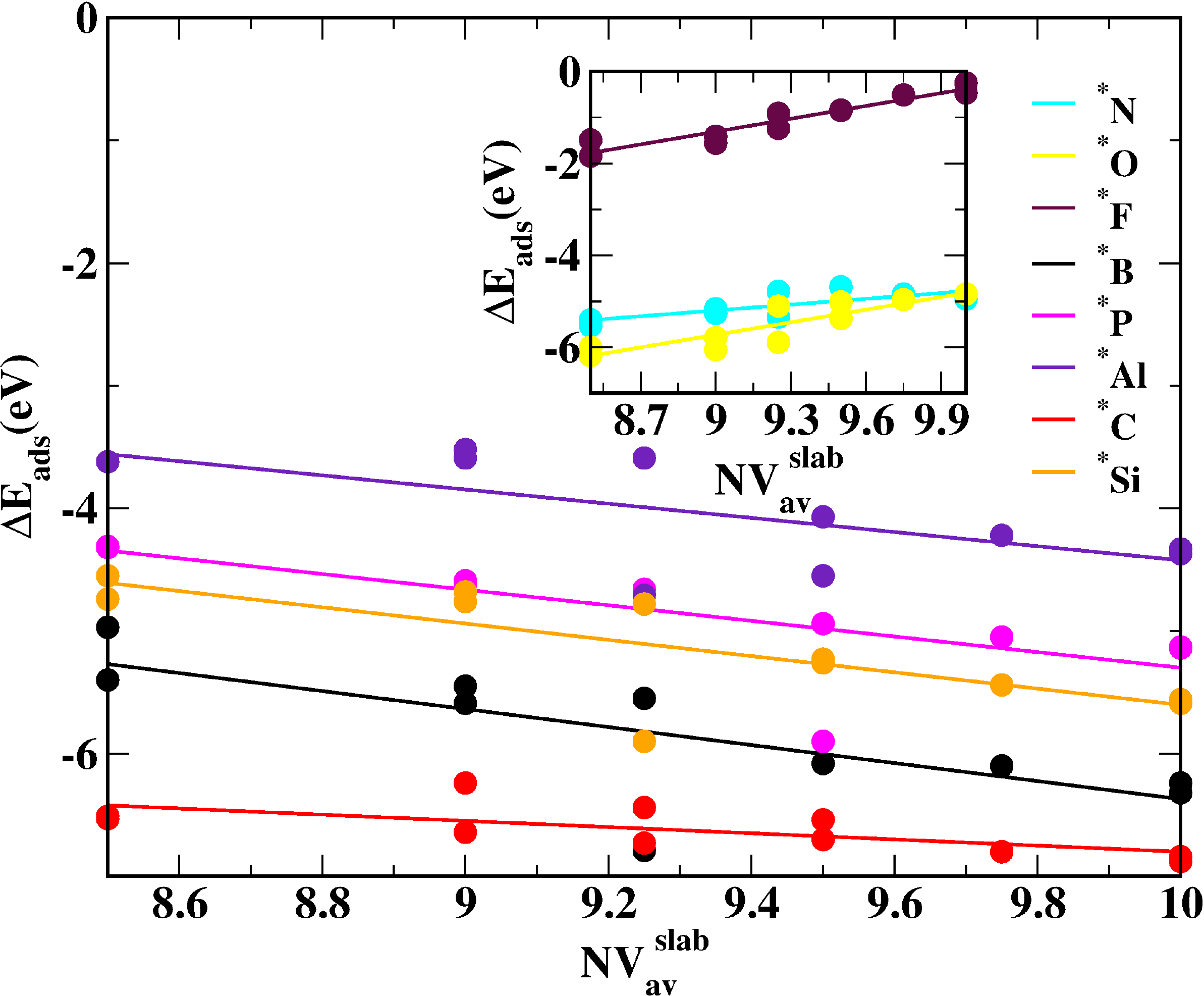}}
     \subfigure[]{\includegraphics[scale=0.29]{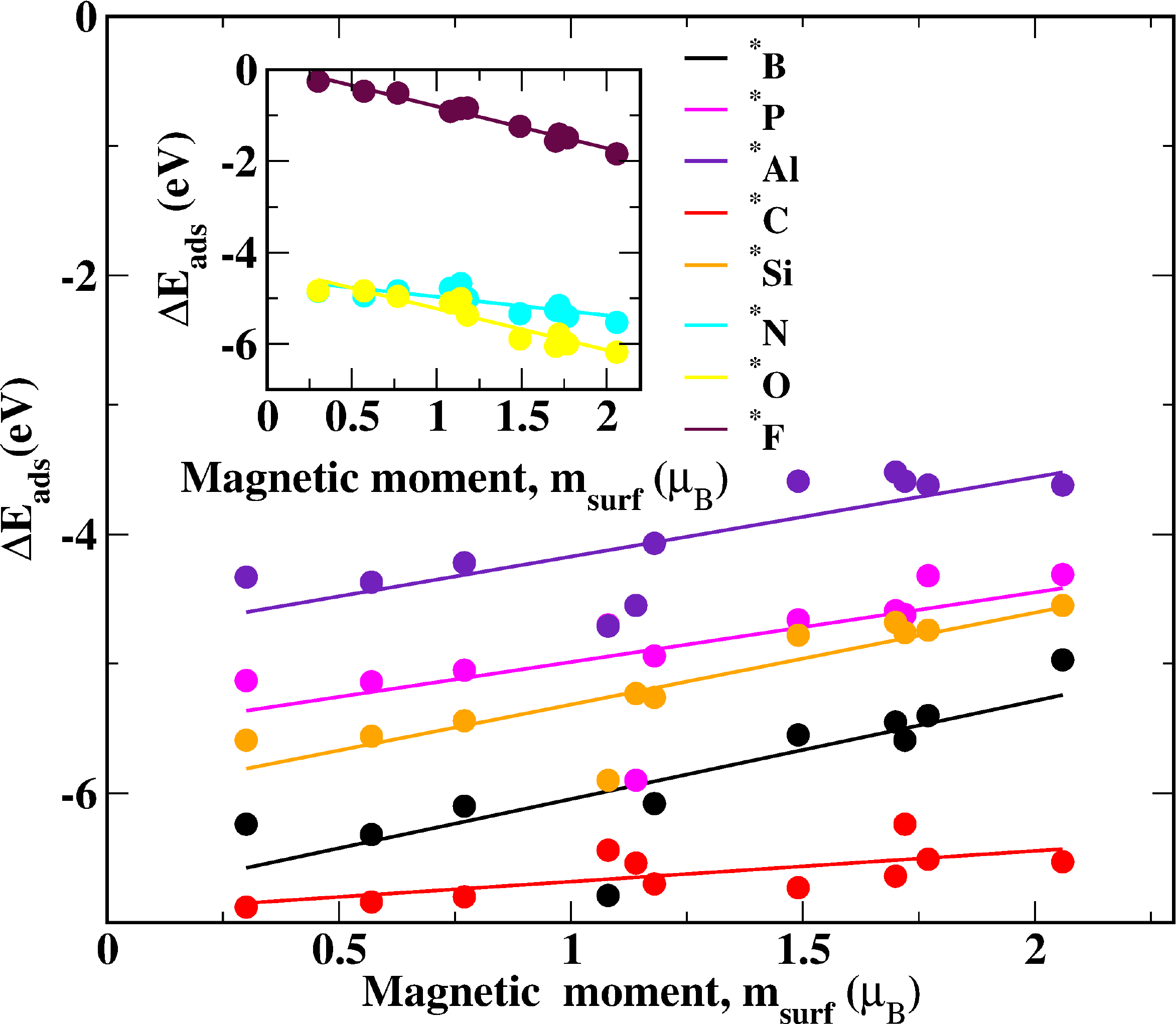}} 
     \subfigure[]{\includegraphics[scale=0.3]{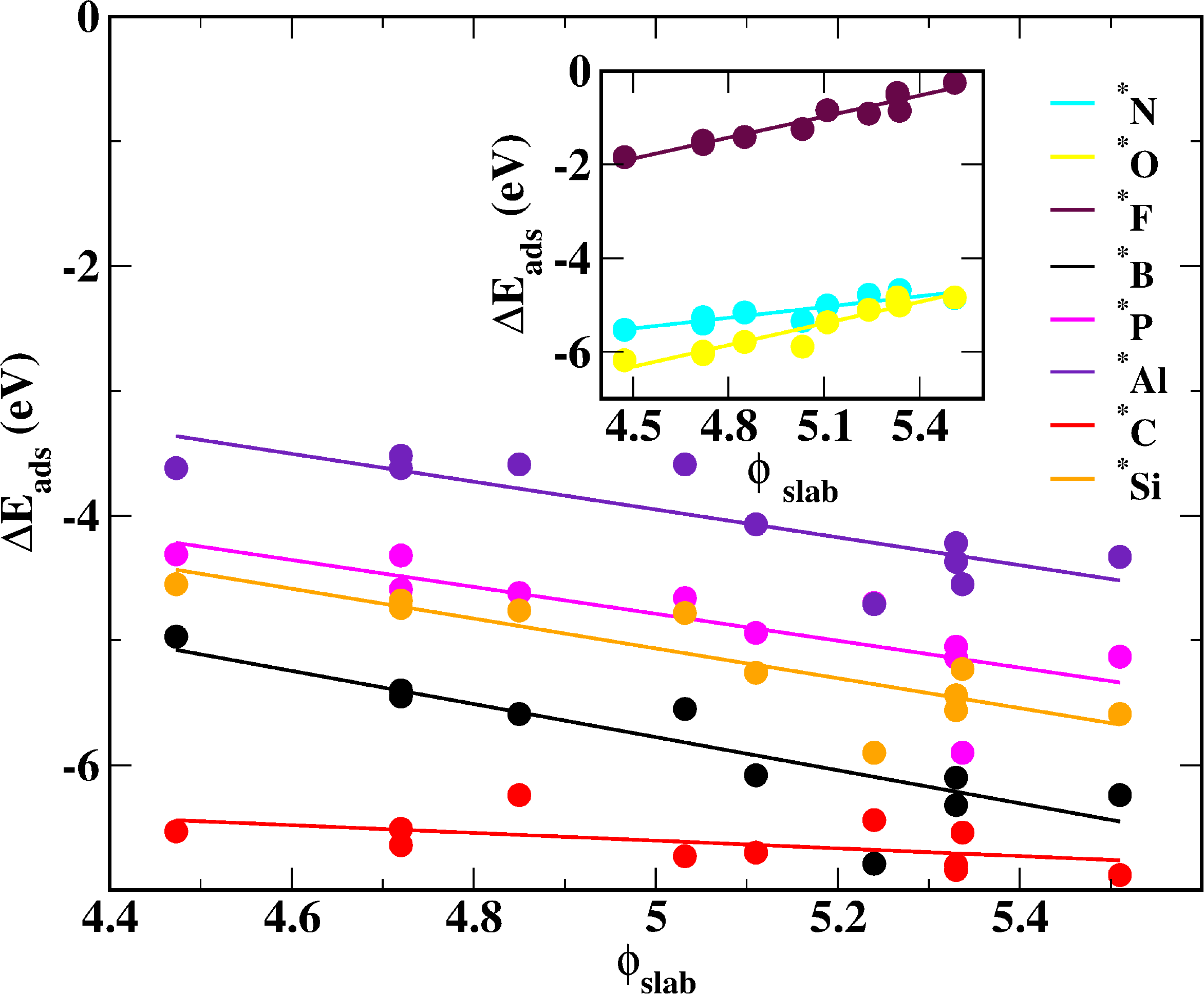}}
     \subfigure[]{\includegraphics[scale=0.3]{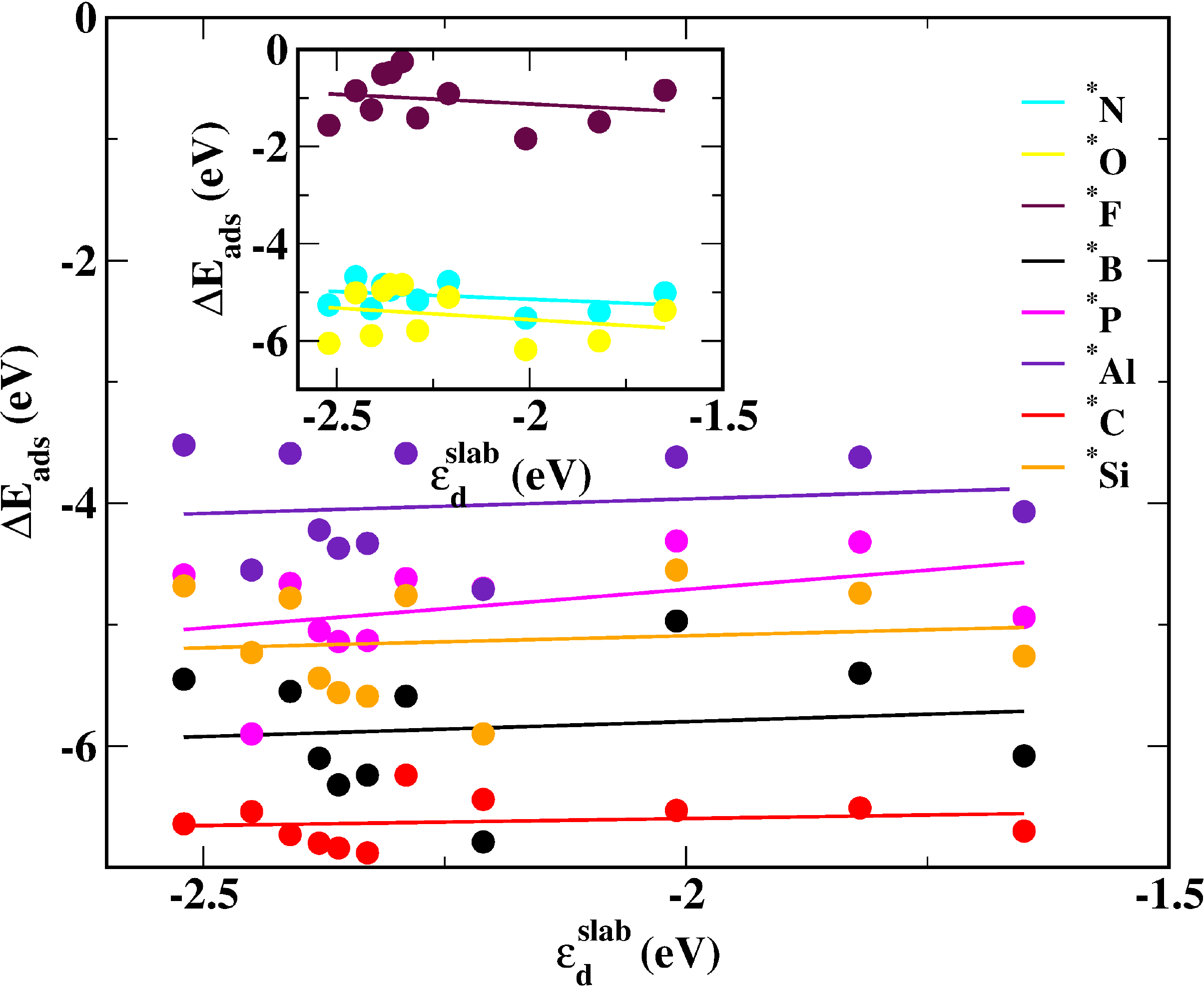}}
        
 \caption{Adsorption energy ($\Delta E_{ads}$) of  B,  C,  N(inset), O(inset), F(inset), Al,  P,  Si as a function of (a) average outer most electrons, $NV_{av}^{slab}$ (b) Magnetic moment of slab, ($m_{surf}$),  (c) work function, $\phi_{slab}$ and (d) spin averaged $d$-band center, $\epsilon_d^{slab}$  for magnetic bimetallic TM surfaces. }
 \label{Fig:scaling-descriptor}
\end{figure}

For further understanding, we have calculated the adsorption energy as a function of different parameters as proposed above and the results are presented in Figure-\ref{Fig:scaling-descriptor}. The occurrence of the linear relation between the proposed parameters and the adsorption energy of $  O$ is noticeable (see Figure-\ref{Fig:scaling-descriptor}). The average outer valence electrons ($NV_{av}^{slab}$) yield a linear scaling relation with negative slope with an adsorption energy of all adsorbates except for $N$, $F$ and $  O$. The decrease (increase) in $\Delta E_{ads}$ energy with increase (decrease) of $NV_{av}^{slab}$ indicating that adsorption gets stronger (weaker). It holds true for other parameters such as $\phi_{slab}$, $m_{surf}$ and $\epsilon_d^{slab}$ as proposed above (see Figure-\ref{Fig:scaling-descriptor}). The related values of equation-\ref{equ:function} are presented in Table-\ref{tab:function}. By comparing the calculated values from Table-\ref{tab-scaling} and Table-\ref{tab:function}, we find that the criterion of equation-\ref{equ:scalingn} and \ref{equ:offset} is accomplished and hence it successfully clarifies the existence of scalability between two adsorbates on magnetic bimetallic  TM surface and simultaneously the appearance of different sign (positive/negative) in the slope of LSR.

From the above observation, we conclude that the scalability is possible in between any two adsorbates and the slope of linear scaling relation is not only depend upon the ratio of the unsaturated bond of adsorbates but also depend on the surface property. Hence, 
 the existence of negative slope in between O and C,  B,  Al,  Si,  P and simultaneously existence of positive slope in between O and N, F is cleared and we propose the equation-\ref{equ:scaling} as 
\begin{equation}
\Delta E_{ads}^{1} = \gamma (m_{surf}, \epsilon_d^{slab}, \phi_{slab}, NV_{av}^{slab} ) \Delta E_{ads}^{2} +\xi \\
\label{equ:scaling2}
\end{equation} 
instead of, 
$\Delta E_{ads}^{1} = \gamma (x_{max}, x) \Delta E_{ads}^{2} +\xi $ as reported originally by Abild-Peterson \textit{et. al} \cite{abild2007scaling}. From the Pearson correlation coefficient matrices (see Figure-S1 in SI), it is cleared that $m_{surf}$ is highly correlated with the other parameters ($\phi_{slab}$ and $NV_{av}^{slab}$), but not with $\epsilon_d^{slab}$. Hence $m_{surf}$ and $\epsilon_d^{slab}$ can be considered as independent variables which has a significant role in the estimation of adsorption energy, as we discussed earlier study\cite{ram2020adsorption}. Therefore equation-\ref{equ:scaling2} can be written as

\begin{equation}
\Delta E_{ads}^{1} = \gamma (m_{surf}, \epsilon_d^{slab} ) \Delta E_{ads}^{2} +\xi \\
\label{equ:scaling3}
\end{equation} 

Identification of simple descriptors
from complex electronic properties of materials becomes
tremendously important which can provide a direction in tailoring the
geometry and composition of surface atoms for desired
properties. The d-band center, is such
a descriptor within the theoretical framework  of surface chemisorption\cite{hammer2000theoretical,hummer1995electronic}.

From the mathematical analysis of equation-\ref{equ:scalingn}, we conclude that for the slope of LSR, $\gamma$ to be positive, both $\frac{\delta F}{\delta \omega_1}$ + $\frac{\delta F}{\delta \omega_2}$  and $\frac{\delta G}{\delta \omega_1}$ + $\frac{\delta G}{\delta \omega_2}$ must be positive or negative. In equation-\ref{equ:scalingn}, for $\omega_i$, i= 1, 2, as we have  two major independent descriptor which can able to explain clearly the existence of LSR, $\omega_1$ = magnetic moment of the surface ($m_{surf}$) and $\omega_2$ = spin-averaged $d$ band center of the surface ($\epsilon_d^{slab}$). 
Similarly for $\gamma$ to be negative, $\frac{\delta F}{\delta \omega_1}$ + $\frac{\delta F}{\delta \omega_2}$ should be positive or negative and at the same time $\frac{\delta G}{\delta \omega_1}$ + $\frac{\delta G}{\delta \omega_2}$ should be vice versa. Again to get $\frac{\delta F}{\delta \omega_1}$ + $\frac{\delta F}{\delta \omega_2}$ and  $\frac{\delta G}{\delta \omega_1}$ + $\frac{\delta G}{\delta \omega_2}$ to be positive or negative we have different cases as explained below.


To get $\frac{\delta F}{\delta \omega_1}$ + $\frac{\delta F}{\delta \omega_2}$ and $\frac{\delta G}{\delta \omega_1}$ + $\frac{\delta G}{\delta \omega_2}$ to be positive there are  different conditions
\begin{equation*}
 \frac{\delta F}{\delta \omega_1} > 0 \quad and \quad  \frac{\delta F}{\delta \omega_2} > 0 \quad or \quad
 \end{equation*}
 \begin{equation*}
\frac{\delta F}{\delta \omega_1} < 0 \quad and \quad
\frac{\delta F}{\delta \omega_2} > 0 \quad but \quad 
\left|\frac{\delta F}{\delta \omega_1}\right| <  \left|\frac{\delta F}{\delta \omega_2}\right|
\quad or \quad
\end{equation*}
 \begin{equation*}
\frac{\delta F}{\delta \omega_1} > 0 \quad and \quad
\frac{\delta F}{\delta \omega_2} < 0 \quad but \quad 
\left|\frac{\delta F}{\delta \omega_1}\right| >  \left|\frac{\delta F}{\delta \omega_2}\right|
\quad and \quad 
\quad similarly \quad
\end{equation*}
\begin{equation*}
 \frac{\delta G}{\delta \omega_1} > 0 \quad and \quad  \frac{\delta G}{\delta \omega_2} > 0 \quad or \quad
 \end{equation*}
 \begin{equation*}
\frac{\delta G}{\delta \omega_1} < 0 \quad and \quad
\frac{\delta G}{\delta \omega_2} > 0 \quad but \quad 
\left|\frac{\delta G}{\delta \omega_1}\right| <  \left|\frac{\delta G}{\delta \omega_2}\right|
\quad
 or \quad
\end{equation*}
 \begin{equation*}
\frac{\delta G}{\delta \omega_1} > 0 \quad and \quad
\frac{\delta G}{\delta \omega_2} < 0 \quad
 but \quad 
\left|\frac{\delta G}{\delta \omega_1}\right| >  \left|\frac{\delta G}{\delta \omega_2}\right|
\end{equation*}

To get $\frac{\delta F}{\delta \omega_1}$ + $\frac{\delta F}{\delta \omega_2}$ and $\frac{\delta G}{\delta \omega_1}$ + $\frac{\delta G}{\delta \omega_2}$ to be negative there are  different conditions
\begin{equation*}
 \frac{\delta F}{\delta \omega_1} < 0 \quad and \quad  \frac{\delta F}{\delta \omega_2} < 0 \quad or \quad
 \end{equation*}
 \begin{equation*}
\frac{\delta F}{\delta \omega_1} < 0 \quad and \quad
\frac{\delta F}{\delta \omega_2} > 0 \quad but \quad 
\left|\frac{\delta F}{\delta \omega_1}\right| >  \left|\frac{\delta F}{\delta \omega_2}\right|
\quad
 or \quad
\end{equation*}
 \begin{equation*}
\frac{\delta F}{\delta \omega_1} > 0 \quad and \quad
\frac{\delta F}{\delta \omega_2} < 0 \quad but \quad 
\left|\frac{\delta F}{\delta \omega_1}\right| <  \left|\frac{\delta F}{\delta \omega_2}\right|
\quad and \quad
\quad similarly \quad
\end{equation*}
\begin{equation*}
 \frac{\delta G}{\delta \omega_1} < 0 \quad and \quad  \frac{\delta G}{\delta \omega_2} < 0 \quad or \quad
 \end{equation*}
 \begin{equation*}
\frac{\delta G}{\delta \omega_1} < 0 \quad and \quad
\frac{\delta G}{\delta \omega_2} > 0 \quad but \quad 
\left|\frac{\delta G}{\delta \omega_1}\right| >  \left|\frac{\delta G}{\delta \omega_2}\right|
\quad
 or \quad
\end{equation*}
 \begin{equation*}
 \frac{\delta G}{\delta \omega_1} > 0 \quad and \quad  \frac{\delta G}{\delta \omega_2} < 0 \quad but \quad 
\left|\frac{\delta G}{\delta \omega_1}\right| <  \left|\frac{\delta G}{\delta \omega_2}\right|
\end{equation*}
 
 In our present case for negative slope (O vs B, C, Al, Si and P) it has satisfied the criteria where $\frac{\delta F}{\delta \omega_1}$ $>$ 0, $\frac{\delta F}{\delta \omega_2}$ $>$ 0 and $\frac{\delta G}{\delta \omega_1}$ $<$ 0, $\frac{\delta G}{\delta \omega_2}$ $<$ 0. As a result  $\frac{\delta F}{\delta \omega_1}$ + $\frac{\delta F}{\delta \omega_2}$ $>$ 0 and  $\frac{\delta G}{\delta \omega_1}$ + $\frac{\delta G}{\delta \omega_2}$ $<$ 0 and for positive slope ( O vs N and F), $\frac{\delta F}{\delta \omega_1}$ $<$ 0, $\frac{\delta F}{\delta \omega_2}$ $<$ 0 and $\frac{\delta G}{\delta \omega_1}$ $<$ 0, $\frac{\delta G}{\delta \omega_2}$ $<$ 0 criteria satisfied, resulting  $\frac{\delta F}{\delta \omega_1}$ + $\frac{\delta F}{\delta \omega_2}$ $<$ 0 and  $\frac{\delta G}{\delta \omega_1}$ + $\frac{\delta G}{\delta \omega_2}$ $<$ 0. The numerical values are reported in SI.

Hence, magnetic moment and spin-averaged d-band center of the surface are the most important parameters that can be related to the appearance of negative slope in the scaling relationships . Therefore the importance of magnetism towards the LSR is justified in this present study which is ignored earlier. In general, with a higher (lower) d-band center\cite{hammer2000theoretical, kitchin2004modification} of a metal surface exhibits
stronger (weaker) affinity to adsorbates due to decreased
(increased) filling of adsorbate-metal antibonding states and the adsorbate-surface interaction can be accomplished. This simple principle has proven to be extremely useful in
search of optimal catalytic materials in many chemical and
electrochemical reactions\cite{hammer2006special,norskov2009towards,norskov2011density}. However this increase or decrease of d-band center of a metal site is not always associated with stronger or weaker chemical bonding. Our present evidence shown in Fig.\ref{Fig:scaling-descriptor}(d)) clarify that the appearance of negative slope does not follow the spin averaged d-band center scaling as we have seen with a higher spin-averaged d-band center exhibits weaker affinity to adsorbates resulting lower the adsorption energies. At the same time positive slopes favour the d-band center concept.
There is a similar correlation between adsorbate-interaction energies and
the substrate d-band centre for transition-state energies,  so-called Brønsted-Evans-Polanyi (BEP)
relation\cite{norskov2008nature}. With the above known
scaling relationships,  the adsorption energies of all the involved
species can be predicted. However, the existence of the such relation limits the search of catalysts. Hence, how to break the
limits of such scaling relationships is a bottleneck question. The above discussion and the presented results clarify the anomaly in LSR is mainly due to the presence of magnetic surface.

Interestingly it is noted that for all considered surfaces, the LSR with positive slope appears between the adatoms specifically in the case of strong electronegative (EN) elements like O, N and F ($EN^O$:3.44, $EN^N$:3.04, $EN^F$:3.98) and for others atomic adsorbate with less electronegativity, LSR appears with a negative slope ($EN^{Al}$:1.61, $EN^{Si}$:1.90, $EN^B$:2.04, $EN^P$:2.19,$EN^C$:2.55). To understand further, the scalability of adsorption energy of $O$ with other $2p$ and $3p$ atomic adsorbates, we analyze in detail through electronic structure calculation.

\section{Charge transfer and Density of states (DOS)}
\begin{figure}[]
  \centering
     {\includegraphics[scale=0.5]{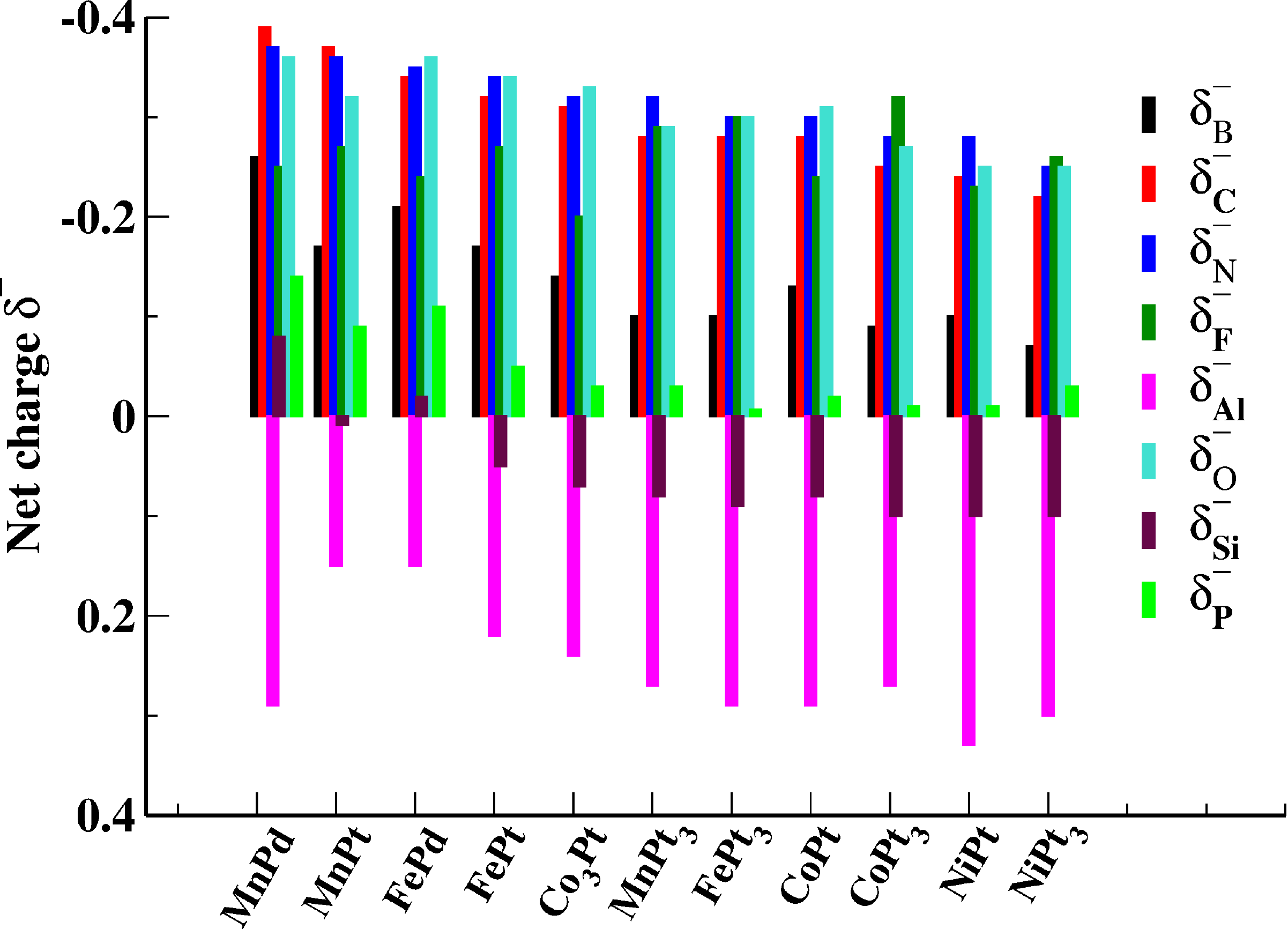}}
    
 \caption{Comparison of charge state of adatoms for magnetic bimetallic  TM surface. 
 The magnitude of net charge transfer for O, N, F and C are comparatively higher for a particular surface than with B, Al, Si and P, which obtain a negative slope for magnetic bimetallic TM surface. At the same time the effective charge on atoms with positive slopes (in the case of  O, N, F) are approximately same and it holds true for negative slope atoms also. The exception is in C case, where the effective charge on C is similar to the O though it has negative slope.}
 \label{Fig:CT}
\end{figure}

Charge transfer is at the basis of
the chemical bond in the ground state. The change in charge transfer sheds a light on why adsorption of atomic adsorbates different. A measure of the localization of charge
around individual atoms is obtained from the calculated
DFT charge density by using Density Derived Electrostatic and Chemical (DDEC6) \cite{limas2016introducing, manz2010chemically} charge analysis method to provide useful insight on charge transfer during an adsorption process. However,
charge transfer is an ambiguous quantity and it is not easy to determine accurately from calculations and experiments. The electron
transfer from the surface to adatoms is different in the
case of positive slope and in negative slope of LSR. The magnitude of excess negative charge on  N and F is nearly equal to the  O adsorbed case in a particular surface which obtain a positive slope on the adsorption process as we can see from the bar diagram in Figure-\ref{Fig:CT} (The calculated values are reported in Table-S4 in SI.). Exceptionally we also find the excess negative charge on  C is approximately same with  O, though the slope is found to be negative in LSR in between C and O (the slope is found to be positive in between O and N(F) in LSR). This may be due to the difference in EN is less in between C and O in comparison to the other element ( $EN^O$:3.44, $EN^B$:2.04, $EN^P$:2.19, $EN^C$:2.55). The detailed analysis for this anomaly in the slope of LSR in  C, we will discuss later. As the effective charge on the O, C, N, F anions becomes more negative in comparison to the B, Al, Si, P, the former anions should interact more strongly with the surface. In addition to this, the earlier study of the scaling relation towards the estimation adsorption energy with a positive slope between any set of adsorbates reflects the fact that the binding of the adsorbates should happen in a similar way with the surface\cite{calle2012physical}. To get more clarification on the above discussion, the electronic structures of the considered systems are
analyzed by studying the projected densities
of states.
 
\begin{figure}[]
  \centering
     \subfigure[]{\includegraphics[scale=0.4]{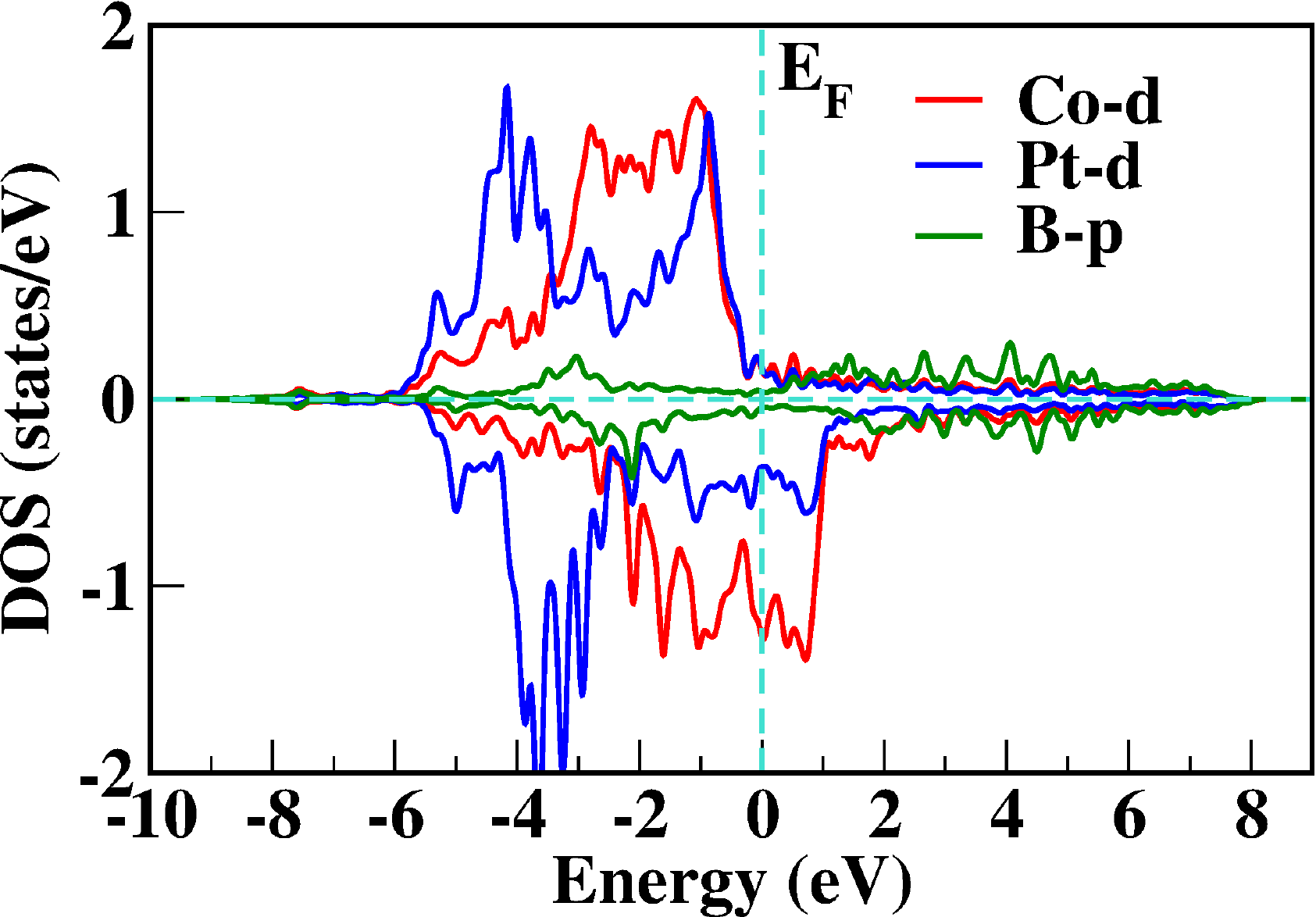}}
     \subfigure[]{\includegraphics[scale=0.4]{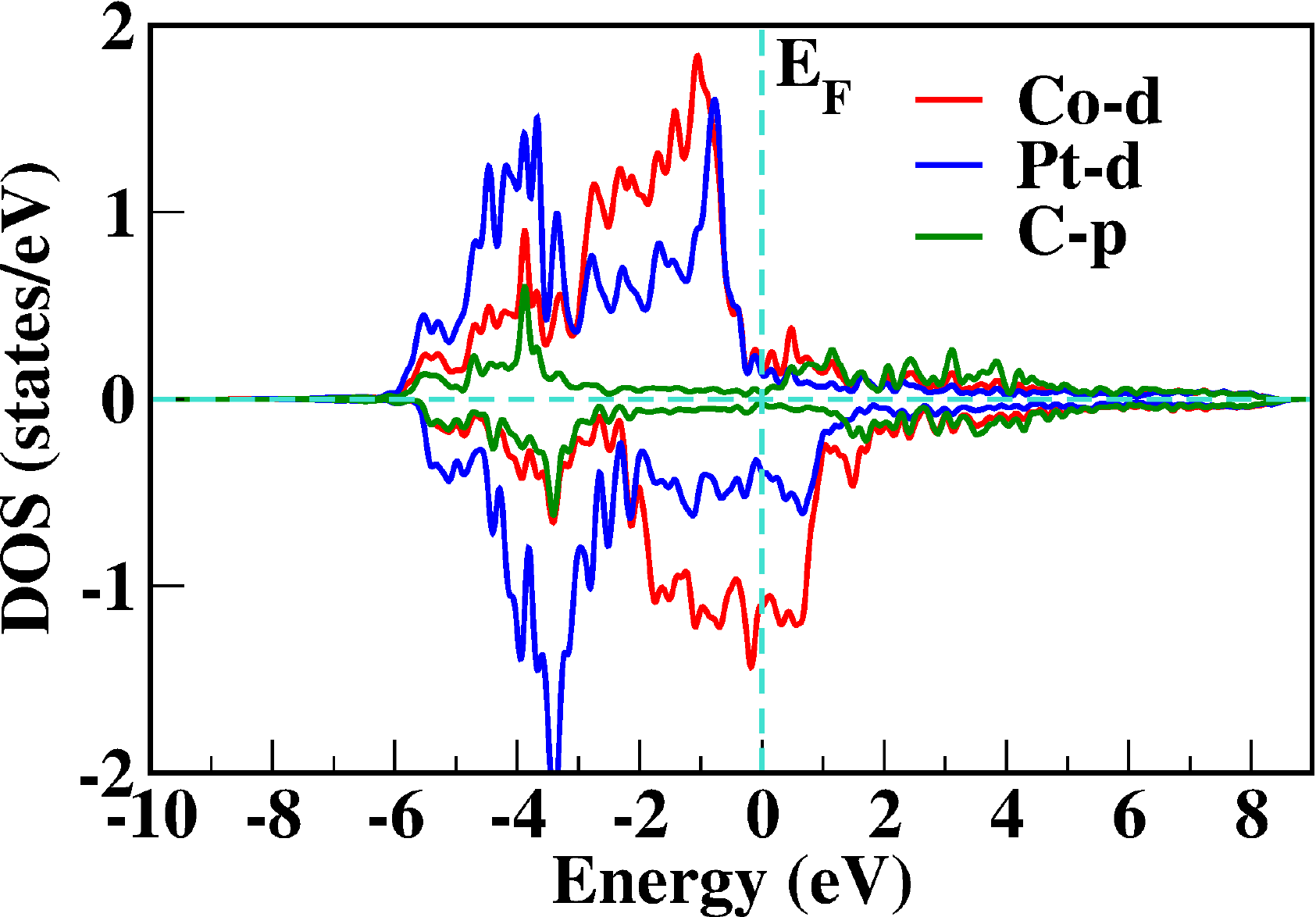}}
     \subfigure[]{\includegraphics[scale=0.4]{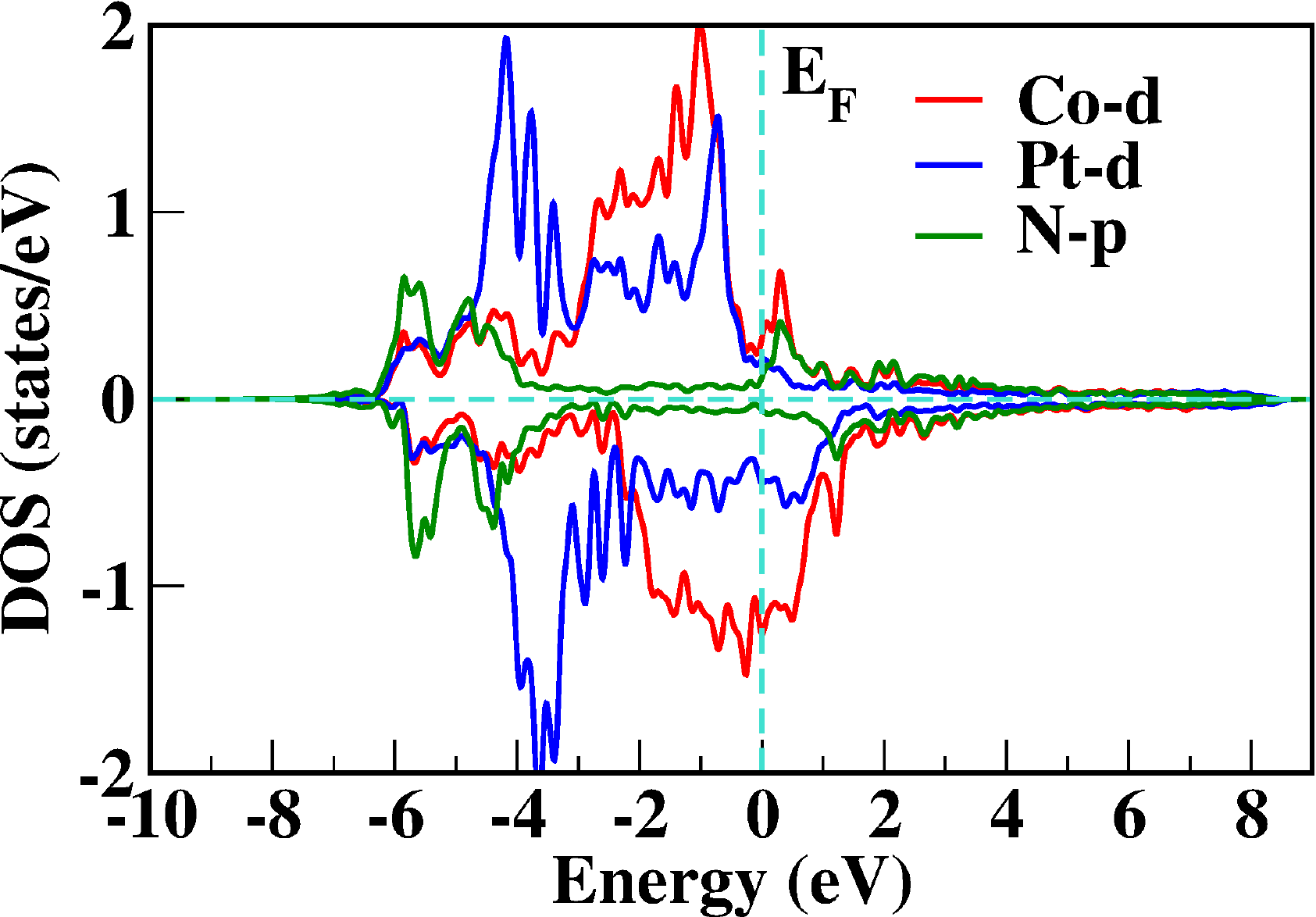}}
     \subfigure[]{\includegraphics[scale=0.4]{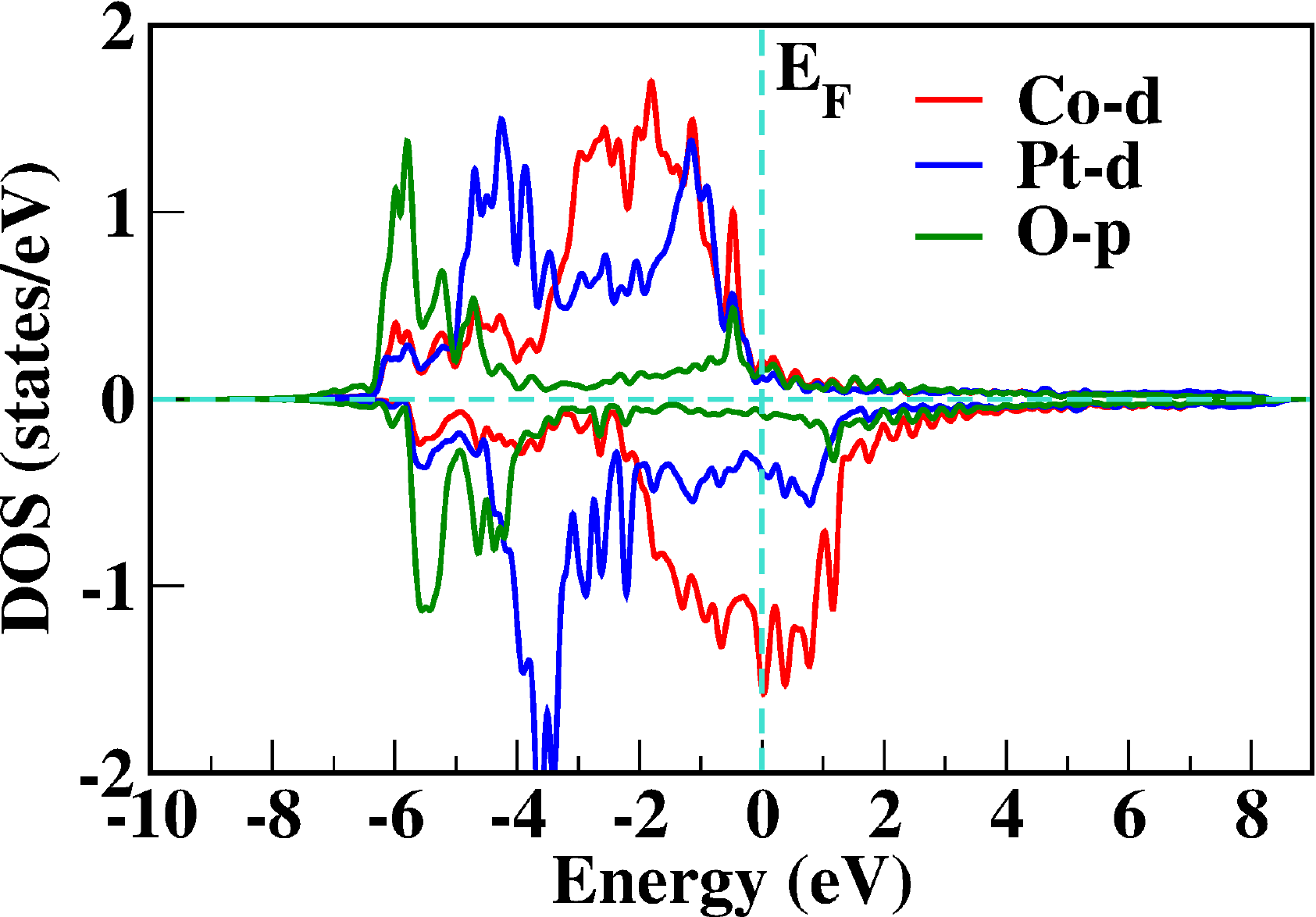}}
     \subfigure[]{\includegraphics[scale=0.4]{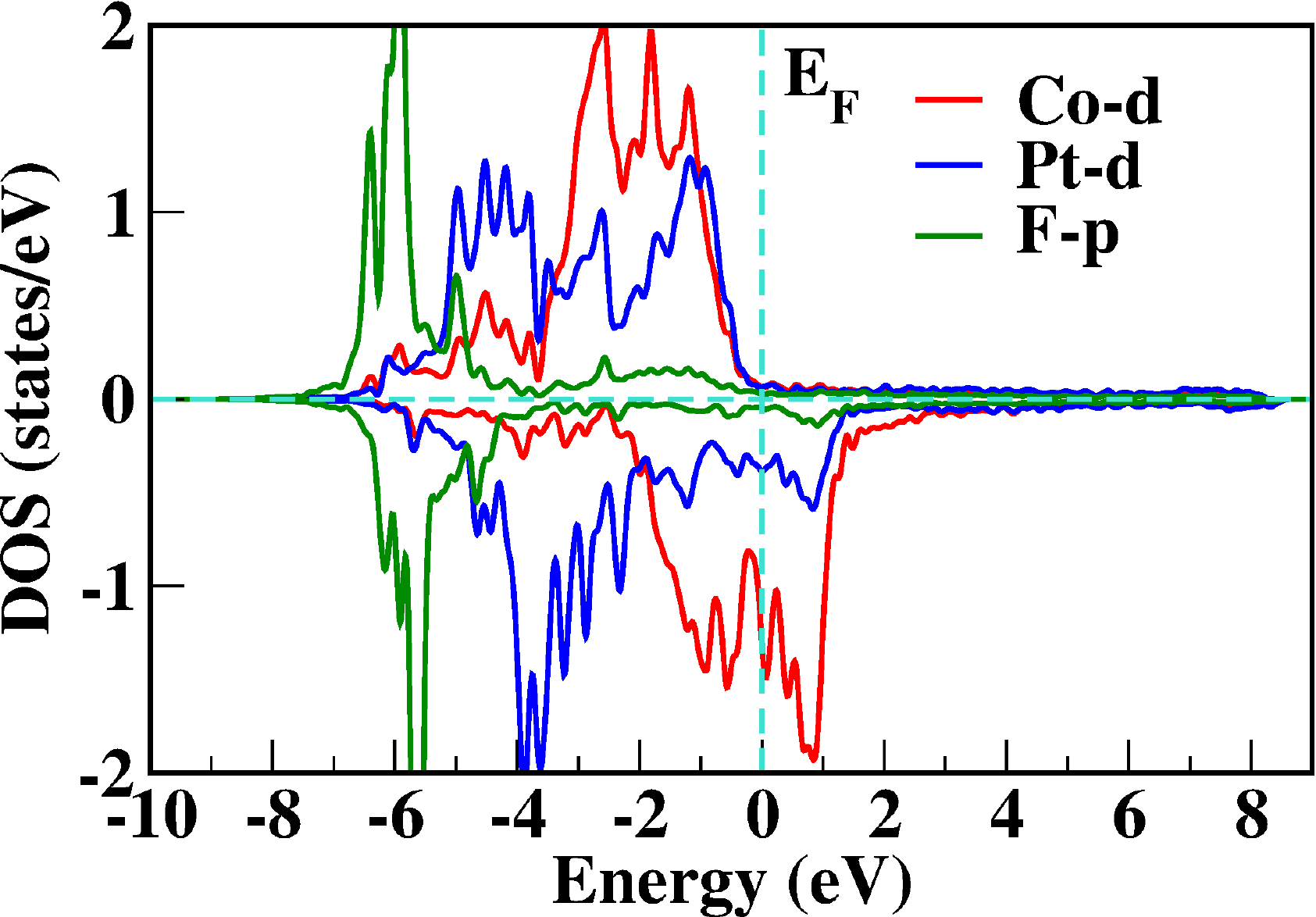}}
     \subfigure[]{\includegraphics[scale=0.4]{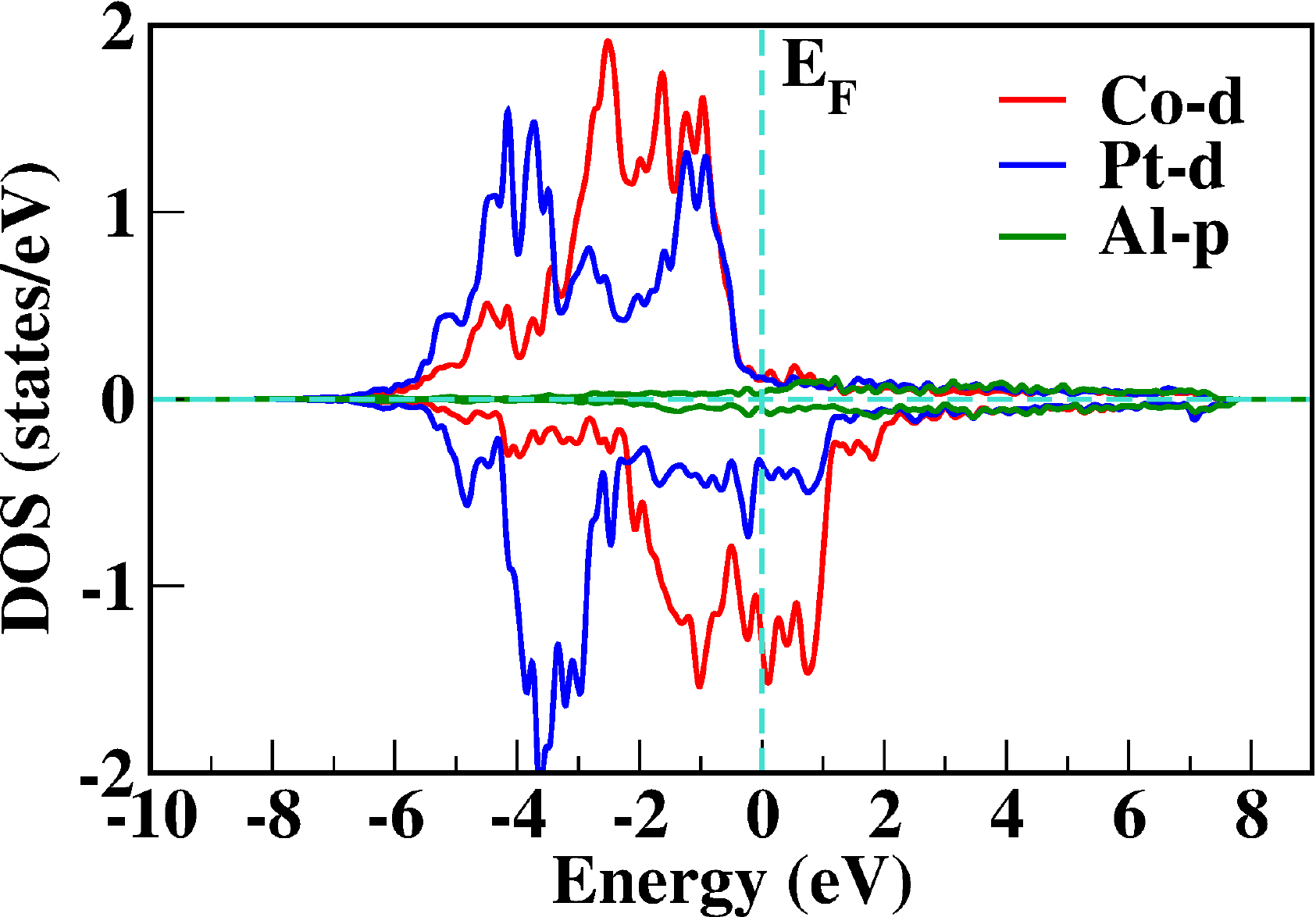}}
     \subfigure[]{\includegraphics[scale=0.4]{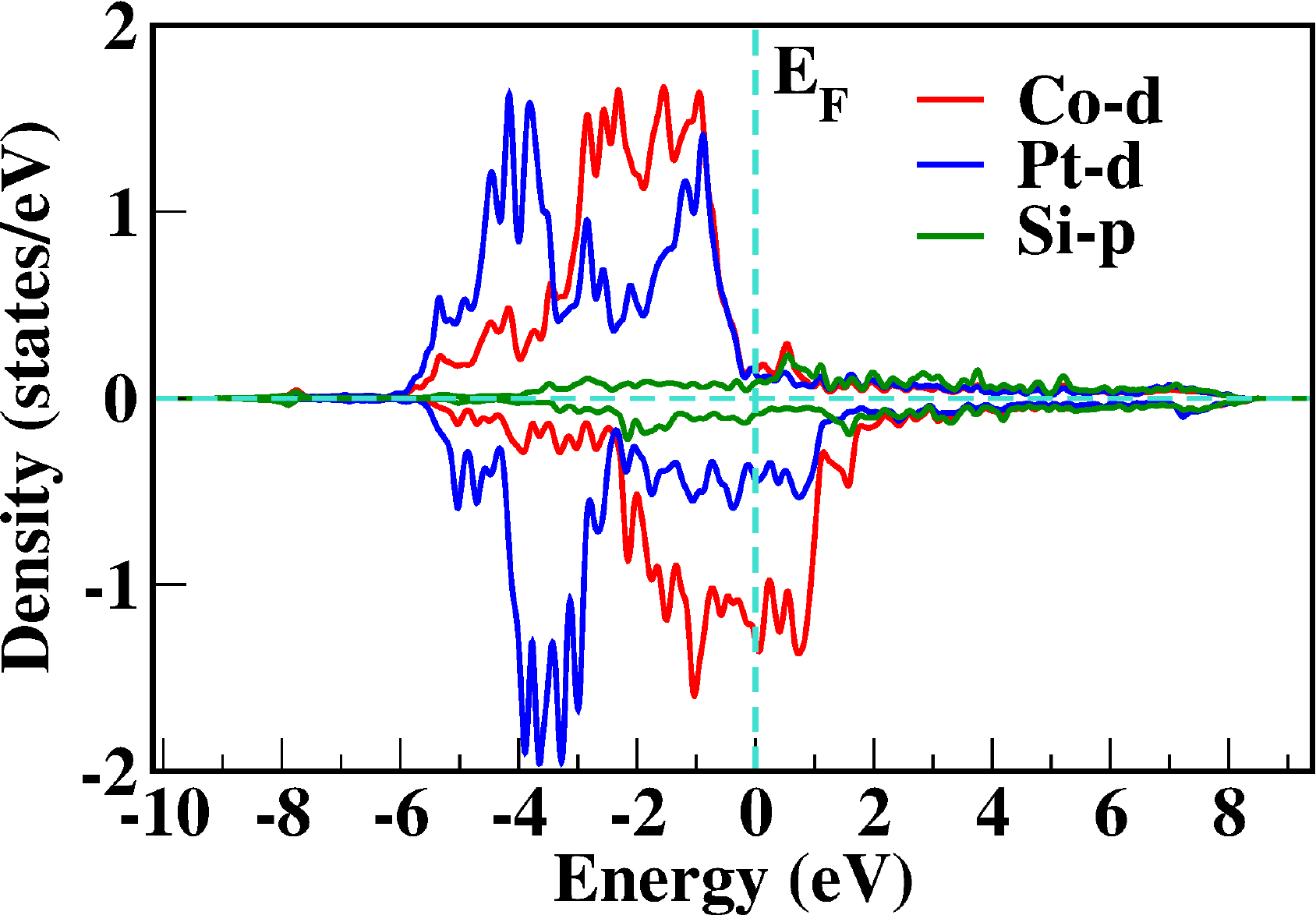}}
     \subfigure[]{\includegraphics[scale=0.4]{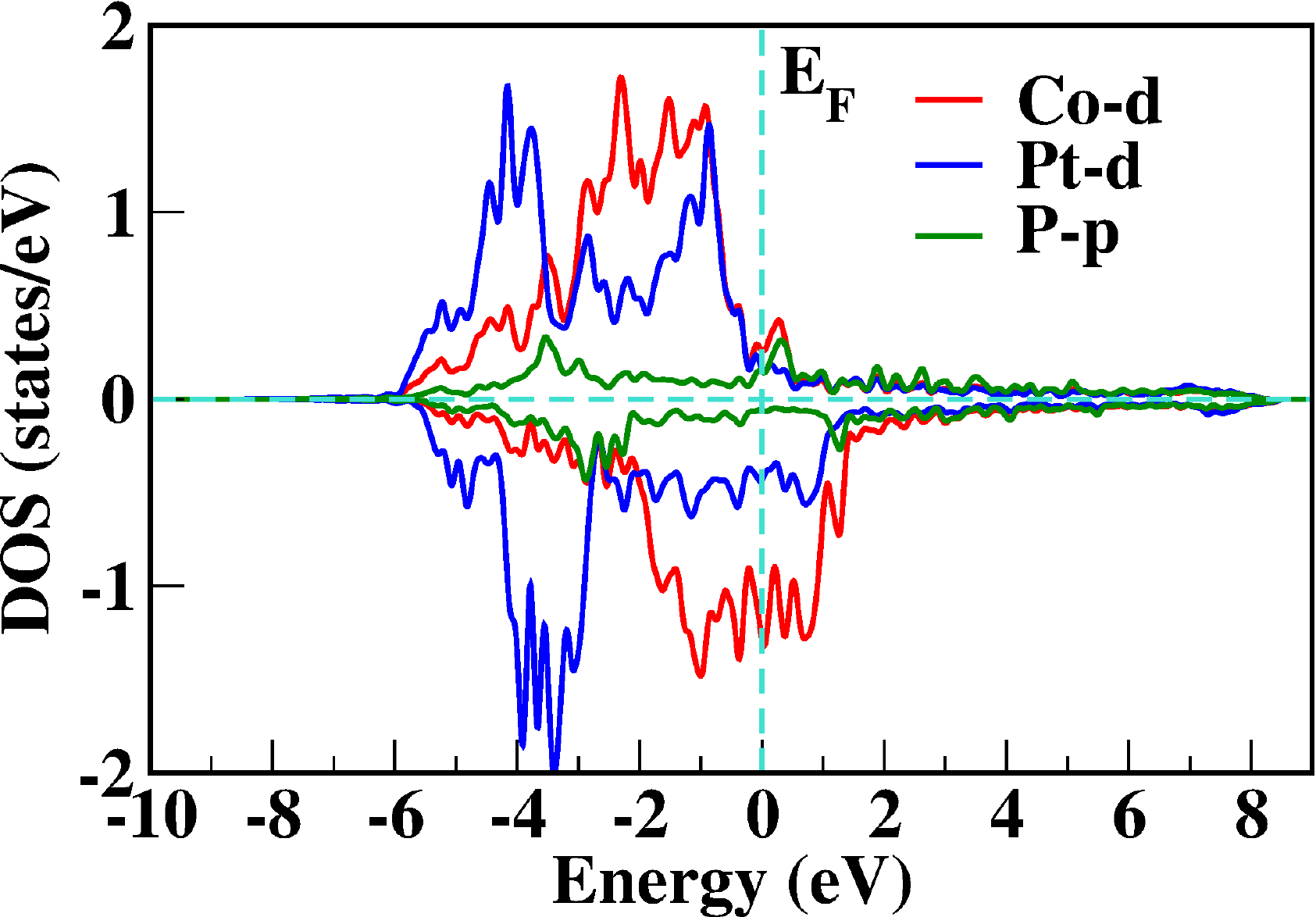}}
   \caption{ Projected density of states of B, C, N, O, F, Al, Si, and P, adsorbed on Co$_3$Pt surface. The scenario is same for other bimetallic surface. The interaction of the $p$ states with surface is observed only in the case of N(F) and O (In the higher energy region -4 eV to -6 eV). }
 \label{Fig:DOS}
\end{figure}

The interaction between the two metal is the main cause for the modification of surface properties in bimetallic systems as a result it changes its electronic environment\cite{liu2001ligand, gauthier2001adsorption}, 
subsequently giving rise to modifications of its electronic structure and consequently, its chemical properties. The magnetic bimetallic TMs are strong ferromagnet with full spin majority bands. As discussed in our earlier paper\cite{ram2020adsorption} the estimated magnetic moment per atom for the magnetic bimetallic TM  surfaces, show an enhanced magnetic moment with respect to its atomic bulk phase. Figure-\ref{Fig:DOS} illustrates that it is the majority spin of Co-$d$ and Pt-$d$ states which strongly hybridized (see an overlap of orbitals only in the majority spin channel in Figure-\ref{Fig:DOS}). This pushes the majority spin states of Co to lower energy, resulting in the enhancement of their occupation and hence in the magnetic moment \cite{bhattacharjee2016improved, bhattacharjee2016nh3,bhattacharjee2015site}. In the other hand, the charge transfer from the Co atoms to Pt is accompanied by a gain in the magnetic moment of Pt. This is evident from the DOS as we can see this transferred electrons asymmetrically occupied the majority and minority band of Pt and induced non-zero magnetic moment at the Pt-site (see Figure-\ref{Fig:DOS}). 

The projected density of states for all adsorbed $2p$ and $3p$ atoms on Co$_3$Pt surface is presented in Figure-\ref{Fig:DOS} (The scenario is the same for other magnetic bimetallic TM surface). The density of states at the Fermi level, $N(E_F)$, has been used in many studies because the electrons at the Fermi level are the most energetically available electrons for the chemical reaction. The value of $N(E_F)$, is however only a single point in the density of states. Furthermore, there has been only limited success in understanding trends in catalysis using $N(E_F)$ (Ref. \cite{van1991theoretical},  Chap.  2). Hammer and  Nørskov\cite{hammer2000theoretical}  have shown that the interactions between adsorbates and transition metal surfaces involve the entire band.
From Figure-\ref{Fig:DOS} we can see the interaction of O atom with the surface atoms of Co$_3$Pt is observed to be in similar way with the interaction of N (Figure-\ref{Fig:DOS}c) and F (Figure-\ref{Fig:DOS}e) case, as we have seen the proper hybridization of $2p$ states of N(F) with the surface atom in lower energy region (nearly -4 eV to -6 eV)). Simultaneously we have not seen the hybridization of $p$ states of other adsorbates (B (Figure-\ref{Fig:DOS}a), C ((Figure-\ref{Fig:DOS}b)), Al (Figure-\ref{Fig:DOS}f), Si (Figure-\ref{Fig:DOS}g), P (Figure-\ref{Fig:DOS}h)) with the surface atoms in the broad energy region. As we have discussed earlier the effective net charge of  O and  C is found to be the same (see Figure-\ref{Fig:CT}), but still we have obtained a negative slope in LSR in the adsorption process for this case. The projected DOS as shown in Figure-\ref{Fig:DOS}(b) clarifies the fact that the hybridization of C-2$p$ states with the surface atoms in Co$_3$Pt is negligible in whole energy range, which ultimately leads to a conclusion that the interaction of surface and C atom to be different (the scenario of same for other selected  magnetic bimetallic surface), resulting negative slope in LSR in the adsorption process. Interestingly the hybridization of $p$ states with the surface is also not seen in the case of  B,  Al,  Si and  P case which also have a negative slope with respect to the  O in the adsorption process. Combining the obtained results as in Figure-\ref{Fig:CT} and Figure-\ref{Fig:DOS}, rationalized the appearance of positive slope in between  O and  N,  F  and of negative slope in between   O and  B,  Al,  Si,  P.

\section{Conclusions}
Despite considerable progress in the theoretical understanding of the linear scaling relation in adsorption energy, the case of magnetic bimetallic TM surfaces are less established than that of pure transition metal surfaces. DFT calculations are performed on the magnetic bimetallic  surface to understand the positive and negative slope in the linear scaling relation in the adsorption process for a selective atomic $2p$ and $3p$ adsorbates. Though the effective net charge on adatoms (O, N, F and C) is found to be the same for a particular surface we obtained a negative slope in LSR in between  O and  C. Apart from this we have also obtained a negative slope in between  O and  B,  Al,  Si,  P. The reason for the appearance of the negative slope is related to two surface descriptors and is related namely spin-averaged d-band center and magnetic moment. Ultimately, the goal is to identify suitable parameters that can be used to reflect the different adsorption trends on these surfaces and hence establishing a linear scaling relation and conclusively helps in designing new materials.

\section{Acknowledgment}
This work was supported by NRF grant funded by MSIP, Korea (No.2009-0082471 and No. 2014R1A2A2A04003865), the Convergence Agenda Program (CAP) of the Korea Research Council of Fundamental Science and Technology (KRCF)and GKP (Global Knowledge Platform) project of the Ministry of Science, ICT and Future Planning.

\bibliography{ADS2}


\newpage

\begin{table}[H]
\caption{Parameters of the scaling relations of Figure-\ref{Fig:scalingOvs2p} related to the Equation-\ref{equ:scaling}.}
\label{tab-scaling}
\begin{tabular}{@{}lll@{}}
\toprule
Scaling & intercept & slope \\ \midrule
B vs O  & -10.449   & -0.84 \\
N vs O  & -2.396    & 0.49  \\ 
C vs O  & -7.56     & -0.17 \\
F vs O  & 3.999     & 0.92  \\
P vs O  & -8.47     & -0.66 \\
Al vs O & -8.14     & -0.76 \\
Si vs O & -9.42     & -0.78 \\ \midrule
\end{tabular}
\end{table}

\begin{table}[]
\caption{The values of all parameters (average outer most electron, $NV_{av}^{slab}$, Magnetic moment of slab, ($m_{surf}$), work function, $\phi_{slab}$ and spin averaged $d$-band center, $\epsilon_d^{slab}$ of slab) related to Equation-\ref{equ:function} for all selected atomic adsorbates.}
\label{tab:function}
\begin{tabular}{@{}llll@{}}
\toprule
adsorbates          & parameters & intercept & slope  \\ \midrule
\multirow{2}{*}{B}  & $N_{av}^{slab}$   & 0.96      & -0.73  \\
                    & $m_{surf}$     & -6.8      & 0.76   \\
                    & $\phi_{slab}$  & 0.85     & -1.33   \\  
                    & $\epsilon_d^{slab}$  & -5.31     & 0.246   \\ \hline

\multirow{2}{*}{C}  & $N_{av}^{slab}$   & -4.27      & -0.25 \\
                    & $m_{surf}$     & -6.92     & 0.24   \\
                    & $\phi_{slab}$  & -5.06     & -0.31   \\
                    & $\epsilon_d^{slab}$  & -6.36     & 0.11   \\  \hline

\multirow{2}{*}{N}  & $N_{av}^{slab}$   & -8.98     & 0.42   \\
                    & $m_{surf}$     & -4.56     & -0.4   \\
                    & $\phi_{slab}$  & -8.96     & 0.77   \\
                    & $\epsilon_d^{slab}$  & -5.58     & -0.32   \\ \hline
                    
\multirow{2}{*}{F}  & $N_{av}^{slab}$   & -9.69      & 0.93   \\
                    & $m_{surf}$     & 0.11     & -0.91  \\
                    & $\phi_{slab}$  & -8.62     & 1.5   \\
                     & $\epsilon_d^{slab}$  & -1.92     & -0.4   \\ \hline

\multirow{2}{*}{Al} & $N_{av}^{slab}$   & 1.347     & -0.58  \\
                    & $m_{surf}$     & -4.78     & 0.61   \\
                    & $\phi_{slab}$  & 1.62     & -1.11   \\ 
                    & $\epsilon_d^{slab}$  & -3.49     & 0.24   \\  \hline
                    
\multirow{2}{*}{Si} & $N_{av}^{slab}$   & 1.03     & -0.66  \\
                    & $m_{surf}$     & -6.02     & 0.71   \\
                    & $\phi_{slab}$  & 0.92     & -1.2   \\ 
                    & $\epsilon_d^{slab}$  & -4.7     & 0.2   \\ \hline
                    
\multirow{2}{*}{P}  & $N_{av}^{slab}$   & 1.07      & -0.64  \\
                    & $m_{surf}$     & -5.5      & 0.54   \\
                    & $\phi_{slab}$  & 0.61     & -1.08   \\
                    & $\epsilon_d^{slab}$  & -3.44     & 0.64   \\  \hline
                    
\multirow{2}{*}{O}  & $N_{av}^{slab}$   & -13.94    & 0.91   \\
                    & $m_{surf}$     & -4.32     & -0.9   \\ 
                    & $\phi_{slab}$  & -13.25     & 1.54   \\ 
                    & $\epsilon_d^{slab}$  & -6.52     & -0.479   \\  \midrule
\end{tabular}
\end{table}
\clearpage

\end{document}


\makeatletter
\renewcommand{\fnum@figure}{\figurename~S\thefigure}
\makeatother

\makeatletter
\renewcommand{\fnum@table}{\tablename~S\thetable}
\makeatother

\section{Details of bulk TM compounds}

We have used $L1_0$ phase of bulk NiPt, FePt, FePd, MnPt, MnPd, CoPt and $L1_2$ phase of CoPt$_3$, MnPt$_3$, FePt$_3$, NiPt$_3$ and Co$_3$Pt for calculating theoretical lattice parameter. The obtained optimized lattice parameters and magnetic moment along with the available experimental and other theoretical values are reported in Table  S\ref{bulk}. We have constructed (111) surface  from the optimized lattice parameters to calculate the adsorption energy of adatoms.

\begin{table}
\centering
\caption{The theoretical lattice parameters for bimetallic TM magnetic compounds in bulk form along with available experimental (in parentheses) and other theoretical values and the calculated magnetic moment for bimetallic TM magnetic compounds with other available theoretical and experimental data (in parentheses).  }
\begin{tabular}{lllll}
Compounds & a in \AA          &  & c in \AA   & Mag. mom in $\mu_B$   \\   \hline
MnPd      & 4.11, (3.82$^b$)                                                                    &  & 3.58, (3.68$^b$)                                                                 & 3.42, 3.78$^n$,(4.4$^n$) \\
          &                                                                                     &  &                                                                                  \\
MnPt      & \begin{tabular}[c]{@{}l@{}}4.09, (4.0$^c$)  \\ 3.89$^f$\end{tabular}                &  & \begin{tabular}[c]{@{}l@{}}3.62, (3.67$^c$)\\ 3.64$^f$\end{tabular}       & 3.59, 4.17$^i$       \\
          &                                                                                     &  &                                                                                  \\
FePd      & 3.84, (3.85$^c$)                                                                    &  & 3.77, (3.71$^c$)                                                                 & 3.25, (3.2$^o$) \\
          &                                                                                     &  &                                                                                  \\
FePt      & \begin{tabular}[c]{@{}l@{}}3.87, (3.852$^a$) \\ 3.838$^a$, 3.88$^f$\end{tabular} &  & \begin{tabular}[c]{@{}l@{}}3.75, (3.713$^a$),\\ 3.739$^a$, 3.78$^f$\end{tabular} & 3.32, 2.93(Fe),0.29(Pt)$^p$ \\
          &                                                                                     &  &                                                                                  \\
Co$_3$Pt     & 3.67, 3.66$^d$                                                                      &  &                                                                                 - & 5.75, 5.66$^l$ \\
          &                                                                                     &  &                                                                                  \\
MnPt$_3$     & \begin{tabular}[c]{@{}l@{}}3.94, 3.91$^e$\\ 3.93$^e$\end{tabular}                   &  &                                                                                 - & 4.2, 4.08$^l$,(4.04$^m$) \\
          &                                                                                     &  &                                                                                  \\
FePt$_3$     & \begin{tabular}[c]{@{}l@{}}3.92, (3.866$^a$)\\  3.88$^g$, 3.91$^a$\end{tabular}    &  &                                                                                 - &4.39, 3.30$^i$, 4.454$^l$\\
          &                                                                                     &  &                                                                                  \\
CoPt      & 3.82, 3.79$^f$                                                                      &  & 3.70$^f$                                                                         & 2.33, 1.6(Co)$^k$,0.3(Pt)$^k$ \\
          &                                                                                     &  &                                                                                  \\
CoPt$_3$     & 3.89,  3.83$^d$                                                                     &  &                                                                                 - &  \begin{tabular}[c]{@{}l@{}}3.27, 2.29$^i$\\(2.43$^j$), 2.652$^l$ \end{tabular} \\
          &                                                                                     &  &                                                                                  \\
NiPt      & 3.85, 3.81$^d$                                                                      &  & 3.63, 3.53$^d$                                                                   & 2.15, 1.035 $^i$ \\
          &                                                                                     &  &                                                                                  \\
NiPt$_3$     & 3.89, 3.89$^h$                                                                       &  &  -     & 0.81 \\ \hline                                                                         
\end{tabular}

\label{bulk}
a\cite{sternik2015dynamical}, b\cite{hori2002crystal},
c\cite{villars1985pearson}, d\cite{pearson1958lattice},
e\cite{hong2008magnetism},f\cite{dannenberg2009surface}, 
g\cite{tobita2010antiferromagnetic},
h\cite{pisanty1990band},
i\cite{dannenberg2010first},
j\cite{menzinger1966magnetic},
k\cite{he1991magneto},l\cite{kumar2009magneto},
m\cite{antonini1969magnetization},
n\cite{jun2012structural},o\cite{moruzzi1993structural},
p\cite{staunton2004temperature}
\end{table}

\begin{table}
\centering
\caption{ Adsorption energy, $\Delta E_{ads}$ for stable site (fcc site) of bimetallic TM magnetic systems for atomic adsorbates considering the second row elements in the periodic table viz. B, C, N, O and F and third row elements in the periodic table viz. Al, Si, and P. }
\begin{tabular}{|l|l|l|l|l|l|l|l|l|}
\hline
 
\multirow{2}{*}{Systems } & 
\multicolumn{8}{c|}{$\Delta E_{ads}$ (eV)}              \\  \cline{2-9}
        & B     & C     & N     & O     & F     & Al    & Si    & P      \\
        \hline
MnPd    & -4.97 & -6.53 & -5.53 & -6.18 & -1.84 & -3.62 & -4.55 & -4.31  \\
MnPt    & -5.4  & -6.5  & -5.4  & -6    & -1.49 & -3.62 & -4.74 & -4.32  \\
FePd    & -5.45 & -6.64 & -5.26 & -6.05 & -1.56 & -3.52 & -4.68 & -4.59  \\
FePt    & -5.59 & -6.24 & -5.16 & -5.79 & -1.41 & -3.59 & -4.76 & -4.62  \\
Co$_3$Pt   & -5.55 & -6.73 & -5.34 & -5.89 & -1.24 & -3.59 & -4.78 & -4.66  \\
MnPt$_3$   & -6.79 & -6.44 & -4.78 & -5.1  & -0.91 & -4.71 & -5.9  & -4.7   \\
FePt$_3$   & -5.9  & -6.54 & -4.68 & -5.01 & -0.85 & -4.55 & -5.23 & -5.9   \\
CoPt    & -6.08 & -6.7  & -5.05 & -5.37 & -0.84 & -4.07 & -5.26 & -4.94  \\
CoPt$_3$   & -6.1  & -6.8  & -4.84 & -4.96 & -0.51 & -4.22 & -5.44 & -5.05  \\
NiPt    & -6.32 & -6.84 & -4.95 & -4.84 & -0.47 & -4.37 & -5.56  & -5.14  \\
NiPt$_3$   & -6.24 & -6.88 & -4.85 & -4.84 & -0.25 & -4.33 & -5.59 & -5.13  \\ \hline
\end{tabular}
\end{table}

\clearpage

\section{Pearson correlation coefficient matrix}
\begin{figure}[H]
  \centering
     {\includegraphics[scale=1.2]{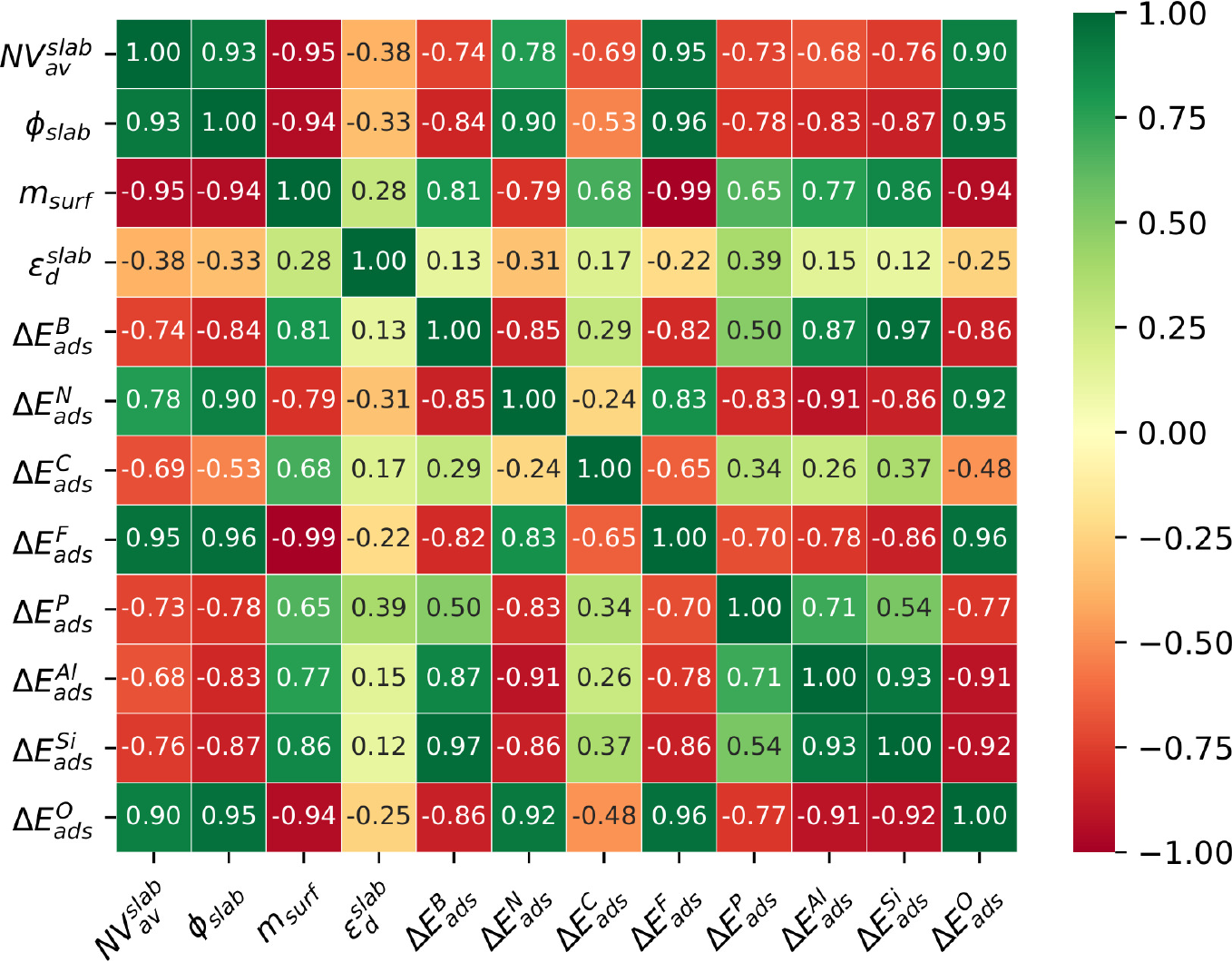}}    
 \caption{Correlation matrix for different parameters viz. average valence electron ($ N_{av}^{slab} $) of slab, work function of clean slab, $ \phi_{slab} $ , magnetic moment projected on surface layers atoms ($m_{surf}$), spin averaged $d$-band center, $\epsilon_d^{slab}$ of slab  and the adsorption energies $\Delta E_{ads}$ of $ B $, $ N $,  $ C $, $ F $, $ P $, $ Al $, $ Si $ and $O $.}
 \label{Fig:correlationmatrix}
\end{figure}

Pearson correlation coefficient matrices (Fig. S\ref{Fig:correlationmatrix}) are calculated from equation-\ref{equ:pearson}, which is a measure of the linear association between two variables $x$ and $y$, where $x, y \in  \lbrace\omega_i\rbrace $  and presented in Fig. S\ref{Fig:correlationmatrix} (see Table S3 in SI).
\begin{equation}
  r (x,y) =
  \frac{ \sum_{i=1}^{n}(x_i-\bar{x})(y_i-\bar{y}) }{%
        \sqrt{\sum_{i=1}^{n}(x_i-\bar{x})^2}\sqrt{\sum_{i=1}^{n}(y_i-\bar{y})^2}}
\label{equ:pearson}        
\end{equation}

\begin{table}[]
\caption{Average valence electron ($ N_{av}^{slab} $) of slab, work function of clean slab, $ \phi_{slab} $ , magnetic moment projected on surface layers atoms ($m_{surf}$), spin averaged $d$-band center, $\epsilon_d^{slab}$ of slab  and the adsorption energies $\Delta E_{ads}$ of $  B $, $  N $,  $  C $, $  P $, $  Al $, $  Si $ and $ O $.}
\begin{tabular}{|llllllllllll|}
\hline
$NV_{av}^{slab}$ & $\phi_{slab}$ & $m_{surf}$ & $\varepsilon_d^{slab}$&$\Delta E^B_{ads}$ & $\Delta E^N_{ads}$ & $\Delta E^C_{ads}$   & $\Delta E^F_{ads}$ & $\Delta E^P_{ads}$     & $\Delta E^{Al}_{ads}$    & $\Delta E^{Si}_{ads}$    & $\Delta E^O_{ads}$ \\
\hline
8.5    & 4.47  & 2.05  & -2.01  & -4.97 & -5.53  & -6.53 & -1.84 & -4.31 & -3.62 & -4.55 & -6.18  \\
8.5    & 4.72  & 1.97  & -1.82  & -5.4  & -5.4  & -6.5 & -1.49  & -4.32 & -3.62 & -4.74 & -6     \\
9      & 4.72  & 1.7   & -2.52  & -5.45 & -5.26 & -6.64 & -1.56 &  -4.59 & -3.52 & -4.68 & -6.05  \\
9      & 4.85  & 1.73  & -2.29  & -5.59 & -5.16 & -6.24 & -1.41 & -4.62 & -3.59 & -4.76 & -5.79  \\
9.25   & 5.032 & 1.52  & -2.41  & -5.55 & -5.34 & -6.73 & -1.24 &-4.66 & -3.59 & -4.78 & -5.89  \\
9.25   & 5.24  & 1.03  & -2.21  & -6.79 & -4.78 & -6.44 & -0.91 & -4.7  & -4.71 & -5.9  & -5.1   \\
9.5    & 5.34  & 1.08  & -2.45  & -5.9  & -4.68 & -6.54 & -0.85 & -5.9  & -4.55 & -5.23 & -5.01  \\
9.5    & 5.11  & 1.2   & -1.65  & -6.08 & -5.01 & -6.7  & -3.27 & -4.94 & -4.07 & -5.26 & -5.37  \\
9.75   & 5.33  & 0.74  & -2.38  & -6.1  & -4.84 & -6.8  & -0.51 & -5.05 & -4.22 & -5.44 & -4.96  \\
10     & 5.33  & 0.54  & -2.36  & -6.32 & -4.95 & -6.84 & -0.47 & -5.14 & -4.37 & -5.56  & -4.84  \\
10     & 5.51  & 0.29  & -2.33  & -6.24 & -4.85 & -6.88 & -0.25 & -5.13 & -4.33 & -5.59 & -4.84 \\ 
\hline
\end{tabular}
\end{table}

\subsection{Mathematical analysis of $F(\omega_i)$ and $G(\omega_i)$}
In the present case for negative slope the differential values are as below, which is reported in Table-2.
 
 O vs B
\begin{equation*}
\frac{\delta F^B}{\delta m_{surf}} = 0.76, \frac{\delta F^B}{\delta \varepsilon_d^{slab}} = 0.246     
\end{equation*} 
 \begin{equation*}
\frac{\delta G^O}{\delta m_{surf}} = - 0.9, \frac{\delta G^O}{\delta \varepsilon_d^{slab}} = -0.48     
\end{equation*} 

O vs C
\begin{equation*}
\frac{\delta F^C}{\delta m_{surf}} = 0.24, \frac{\delta F^C}{\delta \varepsilon_d^{slab}} = 0.11     
\end{equation*} 
 \begin{equation*}
\frac{\delta G^O}{\delta m_{surf}} = - 0.9, \frac{\delta G^O}{\delta \varepsilon_d^{slab}} = -0.48     
\end{equation*}

O vs Al
\begin{equation*}
\frac{\delta F^{Al}}{\delta m_{surf}} = 0.61, \frac{\delta F^{Al}}{\delta \varepsilon_d^{slab}} = 0.24     
\end{equation*} 
 \begin{equation*}
\frac{\delta G^O}{\delta m_{surf}} = - 0.9, \frac{\delta G^O}{\delta \varepsilon_d^{slab}} = -0.48     
\end{equation*}

O vs Si
\begin{equation*}
\frac{\delta F^{Si}}{\delta m_{surf}} = 0.71, \frac{\delta F^{Si}}{\delta \varepsilon_d^{slab}} = 0.2     
\end{equation*} 
 \begin{equation*}
\frac{\delta G^O}{\delta m_{surf}} = - 0.9, \frac{\delta G^O}{\delta \varepsilon_d^{slab}} = -0.48     
\end{equation*} 

O vs P
\begin{equation*}
\frac{\delta F^P}{\delta m_{surf}} = 0.54, \frac{\delta F^P}{\delta \varepsilon_d^{slab}} = 0.64     
\end{equation*} 
 \begin{equation*}
\frac{\delta G^O}{\delta m_{surf}} = - 0.9, \frac{\delta G^O}{\delta \varepsilon_d^{slab}} = -0.48     
\end{equation*} 

Similarly in case of positive slope, the differential values are

O vs N
\begin{equation*}
\frac{\delta F^N}{\delta m_{surf}} = -0.4, \frac{\delta F^N}{\delta \varepsilon_d^{slab}} = -0.32     
\end{equation*} 
 \begin{equation*}
\frac{\delta G^O}{\delta m_{surf}} = - 0.9, \frac{\delta G^O}{\delta \varepsilon_d^{slab}} = -0.48     
\end{equation*} 

O vs F
\begin{equation*}
\frac{\delta F^F}{\delta m_{surf}} = -0.91, \frac{\delta F^F}{\delta \varepsilon_d^{slab}} = -0.4     
\end{equation*} 
 \begin{equation*}
\frac{\delta G^O}{\delta m_{surf}} = - 0.9, \frac{\delta G^O}{\delta \varepsilon_d^{slab}} = -0.48     
\end{equation*}

\begin{table}
\centering
\caption{The net charge,  $\delta^-$ of atomic  adsorbates ( the second row elements in the periodic table viz. B, C, N, O and F and third row elements in the periodic table viz. Al, Si, and P) in adsorption process on bimetallic TM magnetic surface.}
\begin{tabular}{|l|l|l|l|l|l|l|l|l|}
\hline
 
\multirow{2}{*}{Systems } & 
\multicolumn{8}{c|}{Net charge $\delta^-$}              \\  \cline{2-9}
        & B     & C     & N     & O     & F     & Al    & Si    & P      \\
        \hline
MnPd    & -0.26 &  -0.39 &  -0.37  & -0.36 &  -0.25  &   0.29  &   -0.08 &    -0.14 \\
MnPt    & -0.26  &   -0.39  &   -0.37   & -0.32     & -0.27 &    0.15 & 0.009  &   -0.09  \\
FePd    & -0.21  &   -0.34  &   -0.35 & -0.36 & -0.24  &   0.15 &  -0.02  &   -0.11  \\
FePt    & -0.17  &   -0.32  &   -0.34  & -0.34 & -0.27  &   0.22 & 0.05  &    -0.05  \\
Co$_3$Pt   & -0.14  &   -0.31  &   -0.32 & -0.33 & -0.2  &  0.24 & 0.07  & -0.03  \\
MnPt$_3$   & -0.1   &   -0.28  &   -0.32 &  -0.29  & -0.29  &   0.27 & 0.08  &  -0.03   \\
FePt$_3$   & -0.1  &    -0.28  &   -0.3 & -0.3 & -0.3  &  0.29 & 0.09  & -0.007   \\
CoPt    & -0.13  &   -0.28  &   -0.3 & -0.31 & -0.24 &    0.29 & 0.08  &    -0.02  \\
CoPt$_3$   & -0.09  &   -0.25  &   -0.28 & -0.27 & -0.32  &   0.27  & 0.1  &  -0.01  \\
NiPt    & -0.1   &   -0.24  &   -0.28 & -0.25  & -0.23  &   0.33 & 0.1   &    -0.01  \\

NiPt$_3$   & -0.07  &   -0.22  &   -0.25 & -0.25 & -0.26  &   0.3 & 0.1  &   -0.03  \\ \hline
\end{tabular}
\end{table}

\newpage

\bibliography{rsc} 
\bibliographystyle{rsc} 